\documentclass[twocolumn,showpacs,preprintnumbers,amsmath,amssymb,report]{revtex4}
\usepackage{graphicx}
\usepackage{dcolumn}
\usepackage{bm}
\newif\ifappendixon
\begin{document}

\preprint{APS/123-QED}

\title{Theory of slow-light solitons}

\author{A.V. Rybin\footnote{http://www.slowlight.org}}

\affiliation{Department of Physics, University of Jyv\"askyl\"a PO
Box 35, FIN-40351 Jyv\"askyl\"a, Finland}
\email{andrei.rybin@phys.jyu.fi}
\author{ I.P. Vadeiko}
\affiliation{School of Physics and Astronomy, University of St
Andrews, North Haugh, St Andrews, KY16 9SS, Scotland}
\email{iv3@st-andrews.ac.uk}

\author{ A. R. Bishop}
\affiliation{Theoretical Division and Center for Nonlinear
Studies, Los Alamos National Laboratory, Los Alamos, New Mexico
87545, USA} \email{arb@lanl.gov}
\date{ January 10, 2005}
\begin{abstract}

In the framework of the nonlinear $\Lambda$-model we investigate
propagation of  solitons in atomic vapors and Bose-Einstein
condensates. We show how the complicated nonlinear interplay between
fast solitons and slow-light solitons in the $\Lambda$-type media
points to the possibility to create optical gates and, thus, to
control the optical transparency of the $\Lambda$-type media. We
provide an exact analytic description of decelerating, stopping and
re-accelerating of slow-light solitons in atomic media in the
nonadiabatic regime. Dynamical control over slow-light solitons is
realized via a controlling field generated by an auxiliary laser.
For a rather general time dependence of the field; we find the
dynamics of the slow-light soliton inside the medium. We provide an
analytical description for the nonlinear dependence of the velocity
of the signal on the controlling field. If the background field is
turned off at some moment of time, the signal stops. We find the
location and shape of the spatially localized memory bit imprinted
into the medium. We discuss physically interesting features of our
solution, which are in a good agreement with recent experiments.

\end{abstract}
\pacs{03.75.Kk, 03.75.Lm, 05.45.-a}
\keywords{Bose-Einstein condensation, optical soliton}
\maketitle

\section{\label{sec1:level1}Introduction.}
Recent progress in experimental techniques for the coherent control
of light-matter interaction opens many opportunities for interesting
practical applications. The experiments are carried out on various
types of materials such as cold sodium atoms \cite{Hau:1999,
Liu:2001}, rubidium atom vapors \cite{Phillips:2001, Bajcsy:2003,
Braje:2003, Mikhailov:2004}, solids \cite{Turukhin:2002,
Bigelow:2003}, photonic crystals \cite{Soljacic:2004}. These
experiments are based on the control over the absorption properties
of the medium and study slow light and superluminal light effects.
The control can be realized in the regime of electromagnetically
induced transparency (EIT), by the coherent population oscillations
or other induced transparency techniques. The use  of each different
materials brings specific advantages important for the practical
realization of the effects. For instance, the cold atoms have
negligible Doppler broadening and small collision rates, which
increases ground-state coherence time. The experiments on rubidium
vapors are carried at room temperatures and this does not require
application of complicated cooling methods. The solids are obviously
one of the strongest candidates for realization of long-living
optical memory. Photonic crystals provide a broad range of paths to
guide and manipulate the slow light. The interest in the physics of
light propagation in atomic vapors and Bose-Einstein condensates
(BEC) is strongly motivated by the success of research on storage
and retrieval of optical information in these media \cite{Hau:1999,
Liu:2001, Phillips:2001, Kocharovskaya:2001, Bajcsy:2003,
Dutton:2004}.

Even though the linear approach to describing these effects based on
the theory of electromagnetically induced transparency (EIT)
\cite{Harris:1997} is developed in detail \cite{Lukin:2003}, modern
experiments require more complete nonlinear descriptions
\cite{Dutton:2004}. The linear theory of EIT assumes the probe field
to be  much weaker than the controlling field. To allow significant
changes in the initial atomic state due to interaction with the
optical pulse, here we go beyond the limits of  linear theory. In
the adiabatic regime, when the fields change in time very slowly,
approximate analytical solutions \cite{Grobe:1994, Eberly:1995} and
self-consistent solutions \cite{andreev:1998} were found and later
applied in the study of processes of storage and retrieval
\cite{Dey:2003}. Different EIT and self-induced transparency
solitons of nonlinear regime were classified and numerically studied
for their stability \cite{Kozlov:2000}. As it was demonstrated by
Dutton and coauthors \cite{Dutton:2001} strong nonlinearity can
result in interesting new phenomena. Recent experiments and
numerical studies \cite{Matsko:2001, Dutton:2004} have shown that
the adiabatic condition can be relaxed allowing for much more
efficient control over the storage and retrieval of optical
information.

\begin{figure}
\includegraphics[width=60mm]{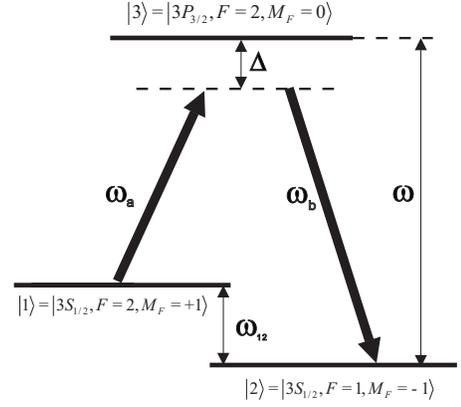}
\caption{\label{fig:spec1} The  $\Lambda$-scheme for working energy
levels of sodium atoms. { The parameters of the scheme are the
following: $\omega_{12}/(2\pi)=1772 \mathrm{MHz}$,
$\omega/(2\pi)=5.1\cdot 10^{14} \mathrm{Hz}$ ($\lambda=589
\mathrm{nm}$), and $\Delta$ is the variable detuning from the
resonance.}}
\end{figure}

In this paper we study the interaction of light with a gaseous
active medium whose working energy levels are well approximated by
the $\Lambda$-scheme. Our theoretical model is a very close
prototype for a gas of sodium atoms, whose interaction with the
light is approximated by the structure of levels of the
$\Lambda$-type. The structure of levels is given in
Fig.~\ref{fig:spec1}, where two hyperfine sub-levels of sodium state
$3^2S_{1/2}$ with $F=1, F=2$ are associated with $|2\rangle$ and
$|1\rangle$ states, { correspondingly~\cite{Hau:1999}}. The excited
state $|3\rangle$ corresponds to the hyperfine sub-level of the term
$3^2P_{3/2}$ with $F=2$. We consider the case when the atoms are
cooled down to microkelvin temperatures in order to suppress the
Doppler shift and increase the coherence life-time for the ground
levels. The atomic coherence life-time in sodium atoms at a
temperature of $0.9 {\mu}$K is of the order  0.9 ms \cite{Liu:2001}.
Typically, in the experiments the pulses have length of
microseconds, which is much shorter than the coherence life-time and
longer than the optical relaxation time of $16.3 ns$.

The gas cell is illuminated by two circularly polarized optical
beams co-propagating in the z-direction. One beam, denoted as
channel $a$, is a $\sigma^-$-polarized field, and the other, denoted
as $b$, is a $\sigma^+$-polarized field. The corresponding fields
are presented within the slow-light varying amplitude and phase
approximation (SVEPA) as
\begin{equation}\label{fields}
     \vec{E}=\vec{e}_a\, \mathcal{E}_a e^{i(k_a z-\omega_a t)} +
     \vec{e}_b\, \mathcal{E}_b e^{i(k_b z-\omega_b t)} +c.c.
\end{equation}
Here, $k_{a,b}$ are the wave numbers, while the vectors
$\vec{e}_a,\vec{e}_b$ describe polarizations of the fields. It is
convenient to introduce two corresponding Rabi frequencies:
\begin{equation}\label{Rabi_f}
    \Omega_a=\frac{2\mu_{a}\mathcal{E}_a}{\hbar},
    \Omega_b=\frac{2\mu_{b}\mathcal{E}_b}{\hbar},
\end{equation}
where $\mu_{a,b}$ are dipole moments of quantum transitions in the
channels $a$ and $b$.

In the interaction picture and within the SVEPA,  the Hamiltonian
$H_\Lambda=H_0+H_I$ describing the interaction of a three-level atom
with the fields is defined as follows:
\begin{eqnarray}
\label{H_lam} H_0=-\frac\Delta2 D,\; H_I=-\frac12 \left({\Omega_a
|3\rangle\langle1| +\Omega_b|3\rangle\langle2|}\right) +h.c.,
\end{eqnarray}
where
$$
D=\left(%
\begin{array}{ccc}
  1 & 0 & 0 \\
  0 & 1 & 0 \\
  0 & 0 & -1 \\
\end{array}%
\right).
$$
Here $\Delta$ is the variable detuning from the resonance and we set
$\hbar=1$.

The dynamics of the fields is described by the Maxwell equations
\begin{eqnarray}\label{Maxwell}
    &(\partial_t^2-c^2 \partial_z^2) \Omega_a e^{i(k_a z-\omega_a
    t)}= -\frac {2 \nu_a}{\omega_a} \partial_t^2 \left({\rho_{31}
    e^{i(k_a z-\omega_a t)}}\right),\nonumber\\
    &(\partial_t^2-c^2 \partial_z^2) \Omega_b e^{i(k_b z-\omega_b
    t)}= -\frac {2 \nu_b}{\omega_b} \partial_t^2 \left({\rho_{32}
    e^{i(k_b z-\omega_b t)}}\right),\nonumber
\end{eqnarray}
where $\nu_a=(n_A |\mu_{a}|^2 \omega_a)/\epsilon_0$, $\nu_b=(n_A
|\mu_{b}|^2 \omega_b)/\epsilon_0$, $n_A$ is the density of atoms,
and $\epsilon_0$ is the vacuum susceptibility. Here $\rho$ is the
density matrix in the interaction representation. For typical
experimental situations  the coupling constants $\nu_{a,b}$ are
almost the same. Therefore we assume that $\nu_a=\nu_b=\nu_0$.
Hence, within the SVEPA the wave equations are reduced to the first
order PDEs:
\begin{eqnarray}\label{Maxwell_r}
    \partial_\zeta \Omega_a = i \nu_0\, \rho_{31}, \;
    \partial_\zeta \Omega_b = i \nu_0\, \rho_{32}.
\end{eqnarray}
Equations Eqs.(\ref{Maxwell_r}) can be rewritten in a matrix form as
\begin{equation}\label{Maxwell_m}
    \partial_\zeta  H_I=i\frac {\nu_0}4 \left[{D, \rho}\right],
\end{equation}
In the new variables the Liouville equation takes the form
\begin{equation}\label{Liouv}
    \partial_\tau \rho=i\left[{\frac\Delta2 D- H_I,\rho}\right].
\end{equation}
Here $\zeta=z/c$, $\tau=t-z/c$. To make parameters dimensionless, we
measure the time in units of optical pulse length $t_p=1\mu s$
typical for the experiments on the slow-light
phenomena~\cite{Bajcsy:2003}. We also normalize the spatial
coordinate to the spatial length of the pulse slowed down in the
medium, i.e. $l_p=v_g t_p\approx c \frac {\Omega_0^2}{2\nu_0}\;t_p$.
Here $\Omega_0$ is a typical magnitude of the controlling  field
required in EIT   experiments. We assume this field to be of order a
few megahertz. We choose $\Omega_0=3$ as a representative value.
This corresponds to a group velocity of several meters per second,
depending on the density of the atoms. We take the group velocity to
be $10^{-7} c$, so the pulse spatial length is $30 \mu m$, and
$\zeta$ is normalized to $10^{-13} s$. Then, in the dimensionless
units, the coupling constant $\nu_0=\frac {\Omega_0^2}2=4.5$. The
retarded time $\tau$ is measured in microseconds and the Rabi
frequencies are normalized to $\mathrm{MH}z$.

The system of equations Eqs.(\ref{Maxwell_m}),(\ref{Liouv}) is
exactly solvable in the framework of the inverse scattering (IS)
method \cite{fad,Park:1998,gab,Rybin:2004}. This means that the
system of equations Eqs.(\ref{Maxwell_m}),(\ref{Liouv}) constitutes
a compatibility condition for a certain linear system, namely
\begin{eqnarray}
 &\partial_\tau \Phi= U(\lambda)\,\Phi=\frac i2\lambda D \,\Phi\,  - i   H_I
 \,\Phi,\label{lin_system_1}\\
& \partial_\zeta \Phi=
V(\lambda)\,\Phi=\frac{i}{2}\frac{\nu_0\rho}{\lambda-\Delta}
\,\Phi\,.\label{lin_system_2}
\end{eqnarray}
Here, $\lambda\in {\Bbb C}$ is the spectral parameter.    The
comparison $\Phi_{\tau\zeta}$ against $\Phi_{\zeta\tau}$ leads to
the zero-curvature condition \cite{fad}
$U_\zeta(\lambda)-V_\tau(\lambda)+\left[U(\lambda),V(\lambda)\right]=0$,
which holds identically with respect to the linearly independent
terms in $\lambda$. It is straightforward to check that the
resulting conditions coincide with the nonlinear equations
Eqs.(\ref{Maxwell_m}),(\ref{Liouv}). At this point it is worth
discussing the initial and boundary conditions underlying the
physical problem in question. We consider a semi-infinite
$\zeta\ge0$ active medium with a pulse of light
 incident at the point $\zeta=0$ (initial condition). This means
 that the evolution is considered with respect to the {\it space}
 variable $\zeta$, while the boundary conditions should be specified
with respect to the  variable $\tau$. In our case we use as the
asymptotic boundary conditions the asymptotic values of the density
matrix at $\tau\to\pm\infty$. To solve the nonlinear dynamics as
described by equations Eqs.(\ref{Maxwell_m}),(\ref{Liouv}), the IS
method considers the scattering problem for the linear system
Eq.(\ref{lin_system_1}), while the auxiliary linear system
Eq.(\ref{lin_system_2}) describes the  evolution of the scattering
data. The purpose of this work is, in particular, to study an
essentially nonlinear interplay of the fields in both the channels.
This goal leads to considering for equation Eq.(\ref{lin_system_1})
the scattering problem of finite density type (cf. \cite{fad} and
references therein), i.e. $\Omega_{a,b}\to\Omega_{a,b}^{\pm}$ as
$\tau\to\pm\infty$. For an account of other results for the
$\Lambda$-system accessible through the IS method see, for example,
references \cite{Hioe:1994, Grobe:1994, Park:1998,gab}. In this work
we choose to use an algebraic version of the IS method, i.e. the
Darboux-B\"acklund (DB) transformations. The DB method does not
require a full investigation of the initial value problem and merely
allows  mounting of a soliton on a chosen background. The resulting
solution is, of course, consistent with the underlying initial value
problem. We use Darboux-B\"acklund transformations, in the spirit of
\cite{ryb1,ryb2, ryb3, salle}, up to certain modifications.

In the physical case considered in this report, the system is
assumed to be initially in a stationary state described by the
following background solution:
\begin{equation}\label{init_fields0}
    \Omega^{}_a=0,\; \Omega^{}_b=\Omega(\tau),\;
    \rho=|\psi_{at}\rangle\langle\psi_{at}|=|1\rangle\langle 1|.
\end{equation}  Notice that the state $|1\rangle$
 is a dark-state for the controlling field $\Omega(\tau)$.
This means that the atoms do not interact with the field
$\Omega(\tau)$ created by the auxiliary laser. The configuration
Eq.(\ref{init_fields0}) above corresponds to a typical experimental
setup (see e.g. \cite{Hau:1999,Liu:2001,Bajcsy:2003}). The function
$\Omega(\tau)$ models the controlling field, which governs the
dynamics of the system. The time dependence of this function can
result from modulation of the intensity of the auxiliary laser. In
general, $\Omega(\tau)$ can also depend on the spatial variable
$\zeta$. However, we do not specify such dependence explicitly in
the formalism below except for a simple case of linear phase shift,
which is discussed in section~\ref{sec3:level1}.

The paper is organized as follows. In the next section we describe
the Darboux-B\"acklund transformation for the $\Lambda$-system. In
section~\ref{sec3:level1} we describe the mechanism  of a
transparency gate for the slow-light soliton. In
section~\ref{sec4:level1} we discuss an exactly solvable example of
manipulation of slow-light solitons, while section~\ref{sec5:level1}
considers a similar problem for the case of a fairly arbitrary
controlling field. Section~\ref{sec6:level1} is devoted to
conclusions and discussion.

\section{\label{sec2:level1} Darboux-B\"acklund transformation for
the $\Lambda$-system}

In this section we describe the Darboux-B\"acklund (DB)
transformation for the $\Lambda$-system. First we reformulate the
linear system Eqs.(\ref{lin_system_1}),(\ref{lin_system_2}) in the
matrix form, viz.

\begin{eqnarray}
 \partial_\tau \Psi&=&\frac{i}{2}D\,\Psi\,{\cal L}  - i  H_I
 \,\Psi,\label{lin_system_1m}\\
 \partial_\zeta \Psi&=&\frac{i\nu_0}{2}{\rho}\,\Psi\,{\cal
P}.\label{lin_system_2m}
\end{eqnarray}
Here $\Psi$ is a matrix consisting of three linearly independent
solutions of the linear system
Eqs.(\ref{lin_system_1}),(\ref{lin_system_2}) corresponding to three
(not necessarily different) values of the spectral parameter
$\lambda$, i.e. $\lambda', \lambda'', \lambda'''$. The matrix
spectral parameters ${\cal L}$  is defined as
$$
{\cal L}=\left(%
\begin{array}{ccc}
  \lambda' & 0 & 0 \\
  0 & \lambda''& 0 \\
  0 & 0 & \lambda''' \\
\end{array}
\right),$$ while ${\cal P}^{-1}={\cal L}-\Delta\cdot I$.

The $N$-fold ($N\ge1$) Darboux-B\"acklund transformation can be
formulated as

\begin{eqnarray}\label{DB_N}
\Psi[N]=\sum\limits_{n=0}^N (-1)^{n+1}\,\Xi_{N-n}(\Delta)\Psi\,
{\cal P}^{-n},\quad \Xi_{0}=I.
\end{eqnarray}

It is clear that the linear system
Eqs.(\ref{lin_system_1m}),(\ref{lin_system_2m}) is covariant with
respect to this transformation provided that  for
 $0\leq n \leq N$ the following Darboux-B\"acklund dressing
 transformations are satisfied

\begin{eqnarray}
 &H_I[N]\,\Xi_{N-n}(\Delta)=\Xi_{N-n}(\Delta)H_I+
 i\partial_\tau\Xi_{N-n}(\Delta) \nonumber\\
\label{Darboux_N}& -\frac12 \left[{D,\left({\Xi_{N-n+1}(\Delta)-
\Delta\,\Xi_{N-n}(\Delta)}\right)}\right]
,\\
\label{Darboux_Na}& \rho[N]\,\Xi_{N-n}(\Delta) = \Xi_{N-n}(\Delta)\,
\rho + \frac{2i}{\nu_0}\partial_\zeta
 \Xi_{N-n+1}(\Delta),\quad
\end{eqnarray}

together with the convention $\Xi_{N+1}(\Delta)\equiv
\Xi_0(\Delta)=I$. The meaning of
Eqs.(\ref{Darboux_N}),(\ref{Darboux_Na}) is that they connect the
"seed" solutions $H_I,\rho$ of the nonlinear system with the dressed
($N$-soliton) solutions $H_I[N],\rho[N]$. To derive the matrices
$\left\{\Xi_k \right\}_{k=1}^N$, we specify a set of solutions
$\left\{\Psi_k\right\}_{k=1}^N$ corresponding to certain fixed
values of the matrix spectral parameter ${\cal L}$, i.e.
$\left\{{\cal L}_k\right\}_{k=1}^N$, where

$${\cal L}_k=\left(%
\begin{array}{ccc}
  \lambda_{k-1}^* & 0 & 0 \\
  0 & \lambda_{k-1}^*& 0 \\
  0 & 0 & \lambda_{k-1} \\
\end{array}%
\right),\; {\cal P}_k^{-1}={\cal L}_k-\Delta\cdot I\,.$$ We then
demand
\begin{eqnarray}\label{Sigma_N}\sum\limits_{n=0}^N (-1)^{n+1}\Xi_{N-n}(\Delta)\Psi_k
\,{\cal P}_k^{-n}=0,\quad k=1,\ldots,N.
\end{eqnarray}
This linear system allows the dressing matrices
$\left\{\Xi_k(\Delta) \right\}_{k=1}^N$ to be obtained through
Cramer's rule. It can be shown that solutions of Eq.(\ref{Sigma_N})
satisfy the relations Eqs.(\ref{Darboux_N}),(\ref{Darboux_Na}).

Since in what follows we only discuss the case $N=1$  for
convenience, we changed notations as follows $H_I[1]\to \tilde H_I$,
$\rho[1]\to \tilde \rho$, $\Xi_1(\Delta)\to\Xi(\Delta)$. Then the
dressing formulae
Eqs.(\ref{DB_N}),(\ref{Darboux_N}),(\ref{Darboux_Na}) reduce to

\begin{eqnarray}\label{Darboux_1}
&\tilde H_I=H_I -\frac12 \left[{D,\Xi(0)}\right],\; \tilde
\rho=\Xi(\Delta)\, \rho\, \Xi^{-1}(\Delta) \\
& \tilde\Psi=\Psi\,{\cal P}^{-1}-\Xi(\Delta)\Psi\label{Darboux_2},
\end{eqnarray}

while from the linear system Eq.(\ref{Sigma_N}) we obtain
\begin{equation}\label{Darboux_2a}
\Xi(\Delta)=\Psi_1({\cal L}_1-\Delta\cdot I)\Psi_1^{-1},
\end{equation}
As was explained above, the matrix $\Psi_1$ is a specification of
$\Psi$ corresponding to a particular value of the matrix spectral
parameter:
$$
{\cal L}_1=\left(%
\begin{array}{ccc}
  \lambda_0^* & 0 & 0 \\
  0 & \lambda_0^*& 0 \\
  0 & 0 & \lambda_0 \\
\end{array}%
\right).
$$

\noindent We denote as $\Phi_0$ the fundamental matrix of solutions
for  the linear system Eqs.(\ref{lin_system_1}),(\ref{lin_system_2})
for $\lambda=\lambda_0$. It can be shown that for the value of the
spectral parameter $\lambda=\lambda_0^*$ the fundamental matrix is
$\bar{\Phi}_0\equiv(\Phi_0^{-1})^\dag$. Since the subspace of
solutions corresponding to $\lambda_0^*$ is two dimensional, the
matrix $\Psi_1$ is constructed as follows. The vector
$\Psi_1^{(3)}=c_1 \Phi_0^{(1)}+c_2 \Phi_0^{(2)}+c_3 \Phi_0^{(3)}$ is
a general solution of the linear problem with $\lambda=\lambda_0$.
Here upper index in the brackets $i=1,2,3$ denotes a vector-column.
To satisfy the structure of the operator $\Xi$ Eq.(\ref{Darboux_2})
we require that $(\Psi_1^{(3)}, \Psi_1^{(1,2)})=0$, where $(\cdot
,\cdot )$ denotes a scalar product of two vectors in 3D complex
vector space, and the vectors $\Psi_1^{(1,2)}$ correspond to
$\lambda=\lambda_0^*$. Due to the definition $(\bar{\Phi}_0^{(i)},
\Phi_0^{(j)})=\delta_{i,j}$.

We can easily find two appropriate orthogonal vectors
$\Psi_1^{(1,2)}$:
$$\Psi_1^{(1)}=
(c_2^*+ c_3^*) \bar{\Phi}_0^{(1)}-c_1^* ( \bar{\Phi}_0^{(2)}+
\bar{\Phi}_0^{(3)});$$
$$ \Psi_1^{(2)}=c_3^* \bar{\Phi}_0^{(2)}-c_2^* \bar{\Phi}_0^{(3)}.$$

The algorithm for finding new solutions of the nonlinear system
Eqs.(\ref{Maxwell_m}),(\ref{Liouv}) can be formulated as follows:
Find a solution $\Phi_0$ of the associated linear system
Eqs.(\ref{lin_system_1}),(\ref{lin_system_2}), corresponding to a
certain "seed" solution of the nonlinear system
Eqs.(\ref{Maxwell_m}),(\ref{Liouv}); Build $\Psi_1$, and build
$\Xi(\Delta)$, then use the dressing transformation
Eq.(\ref{Darboux_1}). It is straightforward to show that for the
state Eq.(\ref{init_fields0}) of the atom-field system  a general
solution of linear system
Eqs.(\ref{lin_system_1}),(\ref{lin_system_2}) can be represented in
the following form
\begin{equation}\label{Gen_solut}
\Phi_0=\left(%
\begin{array}{cc}
  e^{\frac i2\left({\lambda \tau+
  \frac{\nu_0\zeta}{\lambda-\Delta}} \right)} & 0 \\
  0 & \mathrm{T}(\tau,\lambda) \\
\end{array}%
\right),
\end{equation}
where a $2\times2$ matrix $\mathrm{T}(\tau,\lambda)$ is defined
through two complex functions $w(\tau,\lambda)$ and
$z(\tau,\lambda)$ as follows
\begin{eqnarray}\label{Tmatrix}
&    \mathrm{T}(\tau,\lambda)=\left({\mathrm{I}+
\mathrm{W}(\tau,\lambda)}\right)e^{\mathrm{Z}(\tau,\lambda)},\\
&\mathrm{W}(\tau,\lambda)=\left(%
\begin{array}{cc}
  0 & -w^*(\tau,\lambda) \\
  w(\tau,\lambda) & 0 \\
\end{array}%
\right),\nonumber\\
&\mathrm{Z}(\tau,\lambda)=\left(%
\begin{array}{cc}
  i\frac\lambda2\tau +z(\tau,\lambda) & 0 \\
  0 &   -i\frac\lambda2\tau +z^*(\tau,\lambda) \\
\end{array}%
\right).\nonumber
\end{eqnarray}
Here $\mathrm{I}$ is a $2\times2$ identity matrix. The function
$w(\tau,\lambda)$ satisfies the Riccati equation
\begin{equation}\label{Riccati}
    -i\partial_\tau w(\tau,\lambda)=-\lambda w(\tau,\lambda)
    +\frac12 \Omega(\tau)-\frac12 \Omega^*(\tau)\,w^2(\tau,\lambda),
\end{equation}
and the function $z(\tau,\lambda)$ is defined through
$w(\tau,\lambda)$:
\begin{equation}\label{Zequation}
-i\partial_\tau z(\tau,\lambda)=\frac12
\Omega^*(\tau)\,w(\tau,\lambda).
\end{equation}
It is easy to check that $\bar{\Phi}_0$ has the same form as
$\Phi_0$ with $\lambda$ replaced by $\lambda^*$, and $T$ replaced by
\begin{equation}\label{T_bar}
    {\bar{T}}(\tau,\lambda)=\frac{\left({\mathrm{I}+
\mathrm{W}(\tau,\lambda^*)}\right)
e^{-\mathrm{Z}^*(\tau,\lambda^*)}} {1+w(\tau,\lambda^*)\,
w^*(\tau,\lambda^*)}.
\end{equation}

Applying the procedure described above, we find for the fields
\begin{eqnarray}\label{Omega_a_gen}
    &\tilde\Omega_a=-2\, \Xi(0)_{3,1}=\\
    &-2(\lambda_0-\lambda_0^*)e^{-i\,\varphi_1}\frac{
    (w(\tau,\lambda)e^{\varphi_2}+
e^{\varphi_3}) }{\cal{N}},\nonumber \\
\label{Omega_b_gen} &\tilde\Omega_b=\Omega(\tau)-2\,
\Xi(0)_{3,2}=\\
&\Omega(\tau)- 2(\lambda_0-\lambda_0^*)\frac{(e^{\varphi_2^*}-
w(\tau,\lambda^*) e^{\varphi_3^*})
(w(\tau,\lambda)e^{\varphi_2}+e^{\varphi_3}) }{\cal{N}}\,.\nonumber
\end{eqnarray}
The corresponding density matrix
$\tilde\rho=|\tilde\psi_{at}\rangle \langle \tilde \psi_{at}|$
reads
\begin{eqnarray}\label{Wavefun_gen}
&|\tilde\psi_{at}\rangle=\frac{\Xi(\Delta)_{1,1}|1\rangle+
\Xi(\Delta)_{2,1}|2\rangle+ \Xi(\Delta)_{3,1}|3\rangle}
{|\lambda_0-\Delta|}=\\
&\left({\frac{\lambda_0^*-\Delta}{|\lambda_0-\Delta|}+
\frac{(\lambda_0-\lambda_0^*)}{|\lambda_0-\Delta|\cal{N}}}\right)|1\rangle
+\nonumber\\&
\frac{e^{-i\,\varphi_1}(\lambda_0-\lambda_0^*)}{|\lambda_0-\Delta|}
\left({\frac{(e^{\varphi_2}- w^*(\tau,\lambda) e^{\varphi_3}) }
{\cal{N}} |2\rangle+\frac{(w(\tau,\lambda)e^{\varphi_2}+
e^{\varphi_3})}{\cal{N}}|3\rangle}\right).\nonumber
\end{eqnarray}
Here, $\varphi_1$ is the phase of the coefficient $c_1$. The
module of this coefficient can be set to unity without loss of
generality. For shorter notations we have also defined two phases
(cf. Eqs.(\ref{Gen_solut}), (\ref{Tmatrix})):
\begin{eqnarray}\label{phase2}
&\varphi_2=\mathrm{Z}(\tau,\lambda)_{1,1}+\log(c_2)- \frac
i2\left({\lambda \tau+  \frac{\nu_0\zeta}{\lambda-\Delta}}
\right)\nonumber\\
&=z(\tau,\lambda)+\log(c_2)-
\frac{i\,\nu_0\zeta}{2(\lambda-\Delta)},\\
\label{phase3} & \varphi_3=\mathrm{Z}(\tau,\lambda)_{2,2}+\log(c_3)-
  \frac i2\left({\lambda \tau+  \frac{\nu_0\zeta}{\lambda-\Delta}}
\right)\nonumber\\
&=-i\lambda\tau +z^*(\tau,\lambda)+\log(c_3)-
\frac{i\,\nu_0\zeta}{2(\lambda-\Delta)},
\end{eqnarray}
and the normalization function
\begin{eqnarray}\label{N1}
&{\cal{N}}=1+\mathrm{Re}\left[{(w(\tau,\lambda)-w(\tau,\lambda^*))
e^{\varphi_2+\varphi_3^*}}\right]
\nonumber\\&+(1+|w(\tau,\lambda)|^2)( e^{\varphi_2+\varphi_2^*}+
e^{\varphi_3+\varphi_3^*}).\;
\end{eqnarray}
In conclusion of this section, we note an important difference
between $\varphi_2$ and $\varphi_3$. It will be shown below that for
a constant or slowly varying background field the function
$z(\tau,\lambda)$ is of the same order of magnitude as the control
field intensity $|\Omega(\tau)|^2$. Therefore, for small intensities
the phase $\varphi_2$ is slowly varying in time and describes the
slow-light  soliton, while the phase $\varphi_3$ is varying with a
speed close to the speed of light in the vacuum due to the term
$\lambda\,\tau$.

\section{\label{sec3:level1} The transparency gate }

In this section we introduce a concept of fast and slow-light
solitons in the $\Lambda$-medium and explain how the nonlinear
interplay between the solitons leads to  a possibility to control
transparency of the medium. We discuss first the mechanism of
 transparency control  for the slow-light soliton. We explain how the
fast soliton  propagating in the {\it a} channel hops to the {\it b}
channel where the slow-light soliton is propagating. The fast
soliton then destroys the slow-light soliton, thus stopping the
propagation of the latter, and then disappears itself due to the
strong relaxation in the system.

As was indicated above, in this work  we consider exact solutions of
the Maxwell-Bloch system Eqs.(\ref{Maxwell_m}),(\ref{Liouv})
existing on some finite background.  The background field plays the
same role as the controlling field in the conventional linear theory
of EIT, but it enters the exact solutions as a parameter in a
substantially nonlinear fashion. We start with the case of a time
independent field specified as follows
\begin{equation}\label{init_fields1}
    \Omega^{}_a=0,\; \Omega^{}_b=\Omega_0 e^{i k \zeta}.
\end{equation}
Here $k\ll k_{a,b}$ is introduced in order to take into account
small spatial variations of the phase. The intensity of the
background field  $\Omega_0$ is an experimentally adjustable
parameter, which provides  control over the transparency of optical
gates and determines the speed of the slow-light soliton. The
Maxwell-Bloch system Eqs.(\ref{Maxwell_m}),(\ref{Liouv}) is
satisfied with the following initial state of atoms
\begin{equation}\label{atoms_init1}
    \rho_0=\left(%
\begin{array}{ccc}
  1-\frac k{\nu_0} x & 0 & 0 \\
  0 & \frac k{\nu_0} (\frac x2+\Delta) &
  \frac k{\nu_0} \Omega_0 e^{-i k \zeta} \\
  0 & \frac k{\nu_0} \Omega_0 e^{i k \zeta} &
  \frac k{\nu_0} (\frac x2-\Delta) \\
\end{array}%
\right).
\end{equation}
The parameter $x$ determines the population of the excited state and
has to be larger than $2\Delta$. It is important to notice that for
a time-independent background field atoms can be prepared in a
mixture of dark state and polarized states only for nonvanishing
parameter $k$, which allows to access a wider range of physically
interesting situations.

For a time-independent background field we immediately find
solutions of  Eqs.(\ref{Riccati}),(\ref{Zequation}), viz.
\begin{eqnarray}\label{w0_sol}
    &w(\tau,\lambda)=w_0\equiv\frac{\Omega_0 e^{i k
    \zeta}}{\lambda+\sqrt{\lambda^2+{\Omega_0}\!^2}},\\
    \label{w0_sol_1}& z(\tau,\lambda)=z_0\,\tau\equiv
    \frac i2 \Omega_0 e^{-i k \zeta} w_0 \tau= \frac{i\,\Omega_0^2\,\tau}
    {2(\lambda+\sqrt{\lambda^2+{\Omega_0}\!^2})}.
\end{eqnarray}
As we   noted in the previous section, when $\Omega^{}_b$ depends
on $\zeta$
$$
(\Phi_0)_{1,1}= e^{\frac i2\left({\lambda \tau+
  \frac{(\nu_0-kx)\zeta}{\lambda-\Delta}} \right)},
$$
and  the structure of the solution $\mathrm{T}(\tau,\lambda)$ in
Eq.(\ref{Tmatrix}) is slightly modified, i.e. we have to replace
$Z(\tau,\lambda)$ with
\begin{eqnarray}\label{Z1}
\mathrm{Z}_1(\tau,\lambda)=
Z(\tau+\frac{k\zeta}{\lambda-\Delta},\lambda)+\frac{i k x
\zeta}{4(\lambda-\Delta)}I-\frac{i k \zeta}2 \sigma_3,
\end{eqnarray}
where $\sigma_3$ is the Pauli matrix. Hence, the phases
Eqs.(\ref{phase2}), (\ref{phase3}) read
\begin{eqnarray}\label{phase2_k}
&\varphi_2=\log(c_2)+\frac{i\,\Omega_0^2\,\tau}
    {2(\lambda+\sqrt{\lambda^2+{\Omega_0}^2})}- \frac {i\,k\,\zeta}2
\nonumber\\
&+ \frac{i(3kx+2k\sqrt{\lambda^2+{\Omega_0}^2}-2\nu_0)\zeta}
    {4(\lambda-\Delta)},\\
\label{phase3_k} & \varphi_3=\log(c_3)-\frac
{i\,\lambda\,\tau}2-\frac{i\sqrt{\lambda^2+{\Omega_0}^2}\tau}2 +
\frac {i\,k\,\zeta}2
\nonumber\\
&+ \frac{i(3kx-2k\sqrt{\lambda^2+{\Omega_0}^2}-2\nu_0)\zeta}
    {4(\lambda-\Delta)}.
\end{eqnarray}

\noindent Using the general solution
Eqs.(\ref{Omega_a_gen}),(\ref{Omega_b_gen}) we can find the dynamics
describing the formation of the transparency gate for initial
conditions specified in
Eqs.(\ref{init_fields1}),(\ref{atoms_init1}). For simplicity, in
this section we take the spectral parameter to be purely imaginary,
$\lambda_0=i\,\epsilon_0$, and for a solitonic type of solution
$\epsilon_0>\Omega_0$. The solution corresponding to the phases
Eqs.(\ref{phase2_k}), (\ref{phase2_k}) describes the nonlinear
interaction of   fast and slow-light  solitons. This solution is
parameterized by the constants $c_{2,3}$ defining the position and
phase of the two solitons. As we have already indicated above, the
phase $\varphi_2$ determines the position of the slow-light soliton
whereas $\varphi_3$ determines the position of the fast signal. In
practice these constants $c_{2,3}$ are defined by the initial
condition, which specifies the actual pulse of light entering the
medium at the point $z=0$. To understand the structure of the
slow-light soliton one can set $c_3=0$. This choice corresponds to
taking the fast soliton to $-\infty$ in the variable $\tau$. Indeed,
this specification removes the fast pulse component corresponding to
$\varphi_3$ and thus singles out the slow-light soliton part. The
slow-light soliton solution assumes then the following form:
\begin{eqnarray}\label{sl_field_a}
&\tilde\Omega_a=\frac{(\lambda_0^*-\lambda)\,w_0\,
e^{i(\mathrm{Im}\varphi_2-\varphi_1)} }
{\sqrt{1+|w_0|^2}}\mathrm{sech}(\phi_s),
\\ \label{sl_field_b}&\tilde\Omega_b=-\Omega_0\,e^{ik\,\zeta}\,\tanh(\phi_s),
\end{eqnarray}
where
\begin{equation}\label{ss_phase}
\phi_s=\mathrm{Re}\varphi_2+ \frac12\log(1+|w_0|^2)
\end{equation}
is the phase of the slow-light soliton. For simplicity, in the
following we let $k=0$. From the expression above and in the
simplifying approximation
$\frac{\Omega_0^2}{\varepsilon_0^2}<\!\!<1$, $\Delta=0$ the group
velocity of the slow-light soliton can be easily derived:
\begin{equation}\label{ss_vg}
    v_g\approx
c\frac{\Omega_0^2}{\nu_0}.
\end{equation}
The pure state of the atomic subsystem  corresponding to the
slow-light soliton solution reads
\begin{eqnarray}\label{ss_wavefun}
|\tilde\psi_{at}\rangle=&\frac{\mathrm{Re}\lambda-\Delta-i
\mathrm{Im}\lambda\tanh\phi_s} {|\lambda-\Delta|}  |1\rangle -
\frac{\tilde\Omega_a} {2|\lambda-\Delta|w_0} |2\rangle \nonumber\\&
- \frac{\tilde\Omega_a}{2|\lambda-\Delta|} |3\rangle.
\end{eqnarray}
Notice that the population of the upper level $|3\rangle$ is
proportional to the intensity of the background field. The speed
of the slow-light soliton is also proportional to $\Omega_0^2$.
This means that the slower the soliton, the smaller the population
of the level $|3\rangle$  and, therefore, the dynamics of the
nonlinear system as a whole is less affected by the relaxation
process.

To understand the structure of the fast soliton one can choose
$c_2=0$.  We then arrive at an expression describing a signal moving
on   the constant background  with the speed of light (fast
soliton):
\begin{eqnarray}\label{fs_field_a}
&\tilde\Omega_a=\frac{(\lambda_0^*-\lambda)\,
e^{i(\mathrm{Im}\varphi_3-\varphi_1)} }
{\sqrt{1+|w_0|^2}}\mathrm{sech}(\phi_f),\nonumber\\
\label{fs_field_b}&\tilde\Omega_b=-\Omega_0\,e^{ik\,\zeta}\,\tanh(\phi_f),
\end{eqnarray}
where  the phase of the fast soliton is
$$
\phi_f=\mathrm{Re}\varphi_3+ \frac12\log(1+|w_0|^2).
$$
The atomic state is described by the function
\begin{eqnarray}\label{fs_wavefun}
|\tilde\psi_{at}\rangle=&\frac{\mathrm{Re}\lambda-\Delta-i
\mathrm{Im}\lambda\tanh\phi_f} {|\lambda-\Delta|}  |1\rangle +
\frac{w_0^*\tilde\Omega_a} {2|\lambda-\Delta|} |2\rangle
\nonumber\\& - \frac{\tilde\Omega_a}{2|\lambda-\Delta|} |3\rangle.
\end{eqnarray}
We emphasize the principal difference between fast and slow-light
solitons. The slow-light soliton vanishes when the controlling field
is zero due to the factor $w_0$ in Eq.(\ref{sl_field_a}). Another
important feature is that for the slow-light soliton the population
of the upper  level $|3\rangle$ is proportional to $\Omega_0$ and is
small for small background fields and therefore stable to optical
relaxation. In contrast, the amplitude of the fast signal is not
limited by $\Omega_0$ and is determined by the spectral parameter
$\varepsilon_0$. The population of the level $|3\rangle$ is also
defined by $\varepsilon_0$, which means that for large spectral
parameters the fast signal will be attenuated by the relaxation. As
we discuss below, in the absence of the background field the fast
signal behaves as a conventional SIT soliton in a two level system
$|1\rangle\longleftrightarrow |3\rangle$.

\begin{figure}
\includegraphics[width=90mm]{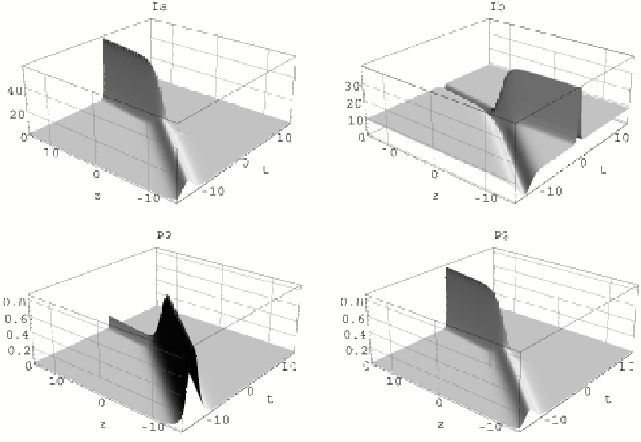}
\caption{\label{fig:spec2} Knocking down the slow-light soliton. The
two upper plots correspond to the dynamics of the intensities of the
fields $\Omega_a$ and $\Omega_b$. The two lower  plots show the
populations of the levels $|2\rangle$ and $|3\rangle$. The
parameters of plotted solutions are $c_2=c_3=1, \lambda_0=4.1i,
\Delta=0$.}
\end{figure}

Figure~\ref{fig:spec2} illustrates the propagation and collision of
the fast and slow-light  solitons according to equation
Eqs.(\ref{Omega_a_gen}), (\ref{Omega_b_gen}). The figure for $I_a$
shows the intensities of the signals in channel $a$. We see that
before the collision only the slow-light soliton exists in channel
$a$, while after the collision the slow-light soliton disappears and
a fast intensive signal appears, whose velocity is slightly below
the speed of light. The figure for $I_a$ is complemented by the
figure for the intensity $I_b$ of the field in channel $b$. The
slow-light soliton corresponds to a groove in the background field
$\Omega_0$. It is clearly seen that after the collision the
slow-light soliton ceases propagating in channel $b$, while some
trace of the fast soliton still can be noticed in that channel. The
process described above can be summarized as if the fast soliton
destroys the slow-light soliton. The notion of a transparency gate
requires the existence of two distinctly different regimes, which
are transparent (open gate), and opaque (closed gate). In the
absence of the fast soliton the gate is open for the slow-light
soliton. When the fast soliton is present the slow-light soliton is
destroyed, while the fast intensive signal created after the
collision in channel $a$ is attenuated due to strong relaxation in
the atomic subsystem. The gate thus closes in the course of the
dynamics due to the relaxation process. To further explain this
process we provide the Fig.~\ref{fig:spec2} plots for populations of
the levels $|2\rangle$ and $|3\rangle$.

Notice that before the collision the population of the upper atom
level $|3\rangle$  is negligible and is approximately given by the
formula for the slow-light soliton solution Eq.(\ref{ss_wavefun})
(see the lower right plot of P3). The populations of the lower
levels $|1,2\rangle$ are determined by the slow-light soliton (see
the lower left plot of P1). Indeed, the fast signal existing in
channel $b$ does not interact with the atoms   because at the onset
of the dynamics their state coincides with the dark state
$|1\rangle$. Figure~\ref{fig:spec2} shows that after the collision
the atoms of the active medium are highly excited and therefore the
level $|3\rangle$ is strongly  populated.  This leads to the fast
attenuation of the rapid intensive signal in channel $a$ due to the
relaxation. The optical gate closes.

To this point we have described a mechanism of controlling  the
transparency of the medium for a particular type of slowly moving
signals. We now  discuss a possibility to read information stored in
the atomic subsystem.   Let us  assume that the background field
vanishes, i.e. $\Omega_0=0$. As was explained above, the speed of
the slow-light soliton then vanishes as well. However, the
information about polarization of the slow-light signal is stored in
the atomic subsystem. This effect can be interpreted in terms of the
concept of a polariton, which is a collective excitation of the
overall atom-field system. The notion of a polariton for the
$\Lambda$-system has been used before. In the linear case the
dark-state polariton was discussed in \cite{Fleischhauer:2000}. In
the strongly nonlinear regime, which is the case for the present
work,  a similar interpretation is possible. Indeed, the field
component of a slow-light soliton solution can be interpreted as the
light contribution into the slow-light polariton. When the
controlling field $\Omega_0$ vanishes this contribution also
vanishes, along with the speed of the polariton. The latter then
contains only excitations in the atomic subsystem. The general
solution Eqs.(\ref{Omega_a_gen}), (\ref{Omega_b_gen}) is then
reduced to the form:

\begin{eqnarray}\label{zfields_field}
 & \tilde\Omega_a=\frac{2i \varepsilon_0
 \exp\left[{i\frac{\Delta\nu_0\zeta}{2(\varepsilon_0^2+\Delta^2)}+
i\log(\frac{c_3}{|c_2|})- i\varphi_1}\right]} {\cosh(\phi_{s0}) +
\frac12 \exp\left[{2\phi_{f0}-\phi_{s0}}\right]},\nonumber\\
  &  \tilde\Omega_b=e^{i\varphi_1+\log(c_2)-\frac{i\nu_0\zeta}
{2(\Delta+i \varepsilon_0)}}\; \tilde\Omega_a,
\end{eqnarray}
where $\phi_{s0}=\frac{\varepsilon_0\nu_0\zeta}
{2(\Delta^2+\varepsilon_0^2)}+ \log(|c_2|)$ is the phase of the
slow-light soliton, and $\phi_{f0}=\varepsilon_0\tau
+\frac{\varepsilon_0\nu_0\zeta}{2(\varepsilon_0^2+\Delta^2)}
+\log(|c_3|)$ is the phase of the fast soliton for the vanishing
background $\Omega_0$. The form of the fields resembles a
superposition of fast and slow-light solitons in
Eqs.(\ref{Omega_a_gen}), (\ref{Omega_b_gen}) with  the vanishing
velocity of the slow-light soliton. This last exponential term in
the denominator represents the fast signal contribution. The
component containing hyperbolic cosine   describes the information
 about the slow-light soliton stored in the
medium after the soliton was stopped. The overall picture of
dynamics corresponds to the scattering of the fast soliton on this
localized atomic polarization pattern. The atomic state describing
this scattering reads
\begin{eqnarray}\label{ss0_wavefun}
  & |\psi\rangle=-
\left({\frac{i\varepsilon_0} {\sqrt{\Delta^2+\varepsilon_0^2}}
\frac{e^{2\phi_{s0}} -1 + e^{2\phi_{f0}}} {e^{2\phi_{s0}}+1 +
e^{2\phi_{f0}}}+\frac{\Delta}{\sqrt{\Delta^2+\varepsilon_0^2}}
}\right) |1\rangle \\
&+\frac{2i\,\varepsilon_0\exp\left[{i\left({\frac{\Delta\nu_0\zeta}
{2(\varepsilon_0^2+\Delta^2)}+\arg(c_2)-\varphi_1}\right)
+\phi_{s0}}\right]}
{\sqrt{\Delta^2+\varepsilon_0^2}\;\left({e^{2\phi_{s0}}+1 +
e^{2\phi_{f0}}}\right)}\;|2\rangle
-\frac{\tilde\Omega_a}{2\sqrt{\Delta^2+ \varepsilon_0^2}}|3\rangle
.\nonumber
\end{eqnarray}
For $c_3=0$ the fields vanish, while the atomic state reduces to a
form corresponding to a stopped polariton described by
Eq.(\ref{ss_wavefun}) with $\Omega_0=0$. In other words, when the
slow-light soliton is completely stopped, the information borne by
the soliton is stored in the spin polarization of the atoms. As
long as the upper state $|3\rangle$ is not populated, the state of
the atomic subsystem is not sensitive to the destructive influence
of the optical relaxation processes.

\begin{figure}
\includegraphics[width=90mm]{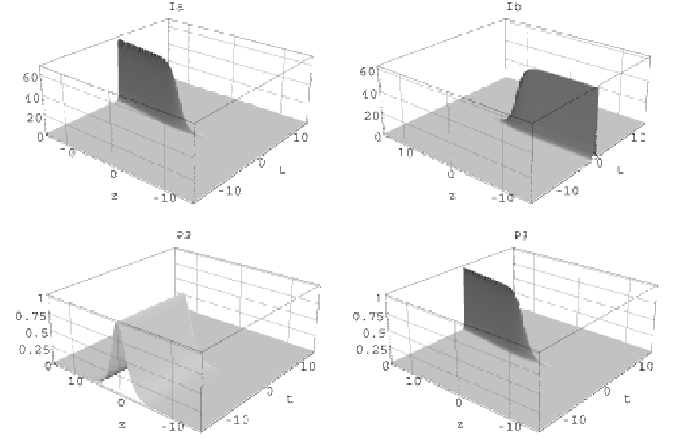}
\caption{\label{fig:spec3} Reading  the optical information by the
fast soliton. The two upper plots illustrate the dynamics  of the
fields $\Omega_a$ and $\Omega_b$. The  two lower  plots show the
populations of the levels $|2\rangle$ and $|3\rangle$. The standing
peak on the plot for $P_2$ corresponds to the stored information in
the form of the  localized polarization. The rapidly moving
localized excitation of the atoms given on the plot for $P_3$
represents the act of reading. The background field $\Omega_0=0$,
and the coupling constant is the same as before, i.e. $\nu_0=4.5$.}
\end{figure}

The  conventional way \cite{Liu:2001} to read the information stored
in the atoms is to  increase the intensity of the background field.
Our method of reading the information is different. We propose to
send the fast soliton into the space domain in the active medium,
where the information is stored.  The polarization in the domain is
then flipped by the fast signal. This is how the reading of
information is realized.  This way of reading optical information is
advantageous  because  it involves fast easily detectable processes.
Figure~\ref{fig:spec3} illustrates  the mechanism of the reading.
Notice that the  act of reading, based on the polarization flipping
induced by the fast signal, can be realized on a very short time
scale compared to typical relaxation times.

\section{\label{sec4:level1}Nonadiabatic manipulation of slow-light solitons}

In this section we discuss an exactly solvable, though physical,
case describing controlled preparation, manipulation and readout of
slow-light solitons in atomic vapors and Bose-Einstein condensates.
 The group velocity of the slow-light soliton depends
explicitly on the field $\Omega_0$, i.e. $v_g\approx
c\frac{\Omega_0^2}{2\nu_0}$ (cf. Eq.(\ref{ss_vg})). This expression
immediately  suggests a plausible conjecture that when the
controlling field is switched off the soliton stops propagating
while the information borne by the soliton remains in the medium in
the form of a spatially localized polarization pattern, i.e. optical
memory, which can be recovered later. For brevity we refer to this
pattern as to "memory bit".

\begin{figure}[htb]
\centerline{\includegraphics[width=70mm]{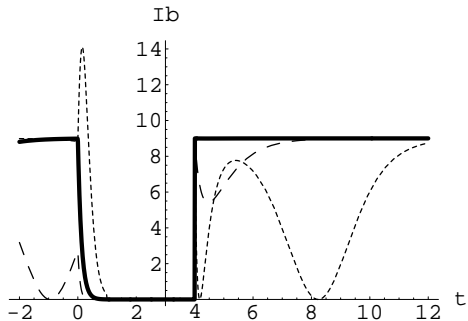}}
\caption{\label{fig1a}  The intensity of the field $\Omega_b$ as the
function of time $t$. The thick solid curve shows the
time-dependence of the controlling background field at entrance into
the medium, i.e. at $z=0$. We also plot $|\tilde\Omega_b|^2$ at
$z=6$ (dashed curve) and at $z=12$ (dotted curve). We choose
$\alpha=4$, and the delay interval is $T-T_1=3$. The dimensionless
units are defined in the text.}
\end{figure}

We consider the following scenario for the dynamics (see
Fig.\ref{fig1a}). Before  $\tau=0$ we create in the medium a
slow-light soliton, and assume it is propagating on the constant
background $\Omega_0$. We then slow-down the soliton by switching
off the laser source of the background field. We assume an
exponential decay of the background field with a decay constant
$\alpha$, i.e. $\Omega_0e^{-\alpha\tau}$. At a certain moment of
time, say, $T_1=4/\alpha$ the field becomes negligible. Therefore,
we cut off the exponential tail and approximate it by zero. At this
step the soliton is completely stopped. The position where the
soliton stops, depends on the decay constant $\alpha$ and on the
moment when we switch  the laser off. The information borne by the
soliton is stored in the form of spatially localized polarization.
This formation can live a relatively long time in atomic vapors or
BEC \cite{Dutton:2004}. At the moment  $T$ we restore the slow-light
soliton by abruptly switching on the laser. The whole dynamics is
divided into four time intervals $\bigcup_{i=0}^3{\cal
D}_i=(-\infty,0]{\cup}[0,T_1]{\cup}(T_1,T]{\cup}(T,\infty)$.
 The time-dependence of the intensity of the background field at
entrance into the medium is given in Fig.\ref{fig1a}.

Before  the soliton enters the medium, the  physical system  is
assumed to be  prepared in the state described by
Eq.(\ref{init_fields0}). The function $\Omega(\tau)$ now models
switching the controlling field off and on again. This function
reads (cf. Fig.\ref{fig1a}):
\begin{eqnarray}
&\Omega(\tau)=
\Omega_0\left[\Theta(-\tau)+e^{-\alpha\tau}(\Theta(\tau)-\Theta(\tau-T_1))\right.\label{exp}\\
&\left.+\Theta(\tau-T)\right]\nonumber.
\end{eqnarray}
Here $\Theta(\cdot)$ is the Heaviside step function with
 $\Theta(0)=\frac{1}{2}$. For the state
Eq.(\ref{init_fields0}) with Eq.(\ref{exp})  we exactly solve the
nonlinear system Eqs.(\ref{Maxwell_m}),(\ref{Liouv}) as well as the
auxiliary scattering problem Eqs.(\ref{lin_system_1}),
(\ref{lin_system_2}) underlying its complete integrability. The
latter result is the cornerstone of analytical progress achieved in
this section. This result allows to mounting a soliton on the
background Eqs.(\ref{init_fields0}),(\ref{exp}) using the
Darboux-B\"acklund transformation described in
section~\ref{sec2:level1} (cf. \cite{Rybin:2004}). According to the
results of section~\ref{sec2:level1} the one-soliton solution
corresponding to the time dependent background
Eq.(\ref{init_fields0}) reads
\begin{eqnarray}\label{fields_tilde}
\tilde\Omega_a&=&\frac{(\lambda^*-\lambda)w(\tau,\lambda)
} {\sqrt{1+|w(\tau,\lambda)|^2}}\; e^{i\tilde\theta_s}\, \mathrm{sech}\tilde\phi_s,\\
\tilde\Omega_b&=&\frac{(\lambda-\lambda^*)w(\tau,\lambda) }
{1+|w(\tau,\lambda)|^2}\,e^{\tilde\phi_s}\,
\mathrm{sech}\tilde\phi_s-\Omega(\tau),\nonumber
\end{eqnarray}
with the atomic state $\tilde\rho=|\tilde\psi_{at}\rangle\langle
\tilde\psi_{at}|$, where
\begin{eqnarray}\label{atoms_tilde}
|\tilde\psi_{at}\rangle=&\frac{\mathrm{Re}\lambda-\Delta-i
\mathrm{Im}\lambda\tanh\tilde\phi_s} {|\lambda-\Delta|}  |1\rangle
\nonumber\\& + \frac{\tilde\Omega_a}
{2|\lambda-\Delta|w(\tau,\lambda)} |2\rangle-
\frac{\tilde\Omega_a}{2|\lambda-\Delta|} |3\rangle.\quad
\end{eqnarray}
Here,
\begin{eqnarray}\label{params_tilde}
\tilde\phi_s&=&\tilde\phi_0+
\frac{\nu_0\zeta}{2}\mathrm{Im}\frac1{\lambda-\Delta}+
\mathrm{Re}(z(\tau,\lambda))\nonumber\\&+&\ln\sqrt{1+|w(\tau,\lambda)|^2},\nonumber\\
\tilde\theta_s&=&\tilde\theta_0-\frac{\nu_0\zeta}2 \mathrm{Re}
\frac1{\lambda-\Delta}+\mathrm{Im}(z(\tau,\lambda)),\nonumber
\end{eqnarray}
\noindent where $\lambda$ is an arbitrary complex parameter. The
functions $w(\tau,\lambda)$, $z(\tau,\lambda)$ are of  piecewise
form, specific to each time region ${\cal D}_i$.
\begin{table*}[htb]
  \centering
  \caption{\label{table1}Exact analytical solution}\begin{tabular}
  {|c|c|c|c|c|}\hline
 ${\cal D}$& $\Omega(\tau)$ & $w(\tau,\lambda)$ & $z(\tau,\lambda)$ & ${\cal C}$
 \\ \hline\hline

${\cal D}_0$ & $\Omega_0$ & $w_0$
   & $\frac i2 \Omega_0 w_0 \tau$
    & 0 \\ \hline

${\cal D}_1$ & $\Omega_0 e^{-\alpha\tau}$ &
    $i\frac{{{\cal C}} J_{1-\gamma}
\left(-\frac{\Omega(\tau)}{2\alpha}\right)- J_{\gamma-1}
\left(-\frac{\Omega(\tau)}{2\alpha}\right)}{{\cal C} J_{-\gamma}
\left(-\frac{\Omega(\tau)}{2\alpha}\right)+ J_{\gamma}
\left(-\frac{\Omega(\tau)}{2\alpha}\right)}$ &
$-\alpha\gamma\tau+\ln \frac{{{\cal C}} J_{-\gamma}
\left(-\frac{\Omega(\tau)}{2\alpha}\right)+ J_{\gamma}
\left(-\frac{\Omega(\tau)}{2\alpha}\right)}{{\cal C} J_{-\gamma}
(-\frac{\Omega_0}{2\alpha})+ J_{\gamma}
\left(-\frac{\Omega_0}{2\alpha}\right)}$ & $\frac{-i w_0
J_{\gamma} (-\frac{\Omega_0}{2\alpha})+ J_{\gamma-1}
(-\frac{\Omega_0}{2\alpha})}{ J_{1-\gamma}
(-\frac{\Omega_0}{2\alpha})+ i w_0 J_{-\gamma}
(-\frac{\Omega_0}{2\alpha})}$ \\\hline

${\cal D}_2$ & $0$ & 0 & $\ln\frac{{\cal C}
\left(-\frac{\Omega_0}{4\alpha}\right)^{-\gamma}/\mathrm{\Gamma}(1-\gamma)}
{{\cal C} J_{-\gamma} (-\frac{\Omega_0}{2\alpha})+ J_{\gamma}
\left(-\frac{\Omega_0}{2\alpha}\right)}$ & ${\cal C}_2={\cal C}_1$
\\\hline

${\cal D}_3$ & $\Omega_0$ & $\frac{\Omega_0
\tan\left({\frac12\sqrt{\lambda^2+\Omega_0^2}(\tau-T)}\right)}
{\lambda\tan\left({\frac12\sqrt{\lambda^2+\Omega_0^2}(\tau-T)}\right)
-i\sqrt{\lambda^2+\Omega_0^2} }$ & $\begin{array}{c}\ln\frac{{\cal
C}\, e^{\frac{-i\left({\lambda+\sqrt{\lambda^2+\Omega_0^2}}\right)
(\tau-T)}2}+ e^{\frac{-i
\left({\lambda-\sqrt{\lambda^2+\Omega_0^2}}\right)
(\tau-T)}2}}{{\cal C}+1}\\+z_2\end{array}$
 & $\frac{\Omega_0^2+2\lambda
\left({\lambda-\sqrt{\lambda^2+\Omega_0^2}}\right)}
{\Omega_0^2}$ \\
\hline
  \end{tabular}
\end{table*}
For clarity we organize elements of the solution corresponding to
different time regions in Table~\ref{table1}. The function $w_0$ is
 defined as in Eq.(\ref{w0_sol}) with $k=0$, the index  of Bessel functions is defined as
$\gamma=({\alpha+i\lambda})/({2\alpha})$. The values ${\cal C}_i $
of the constant ${\cal C}$ for each time region ${\cal D}_i$ are
specified in the rightmost column of the table, the moment of time
$T$  is chosen as in Fig.\ref{fig1a}, i.e. $T=4/\alpha+3$. Notice
that in the table $w_2=w_1(\infty,\lambda)$ and
$z_2=z_1(\infty,\lambda)$. Therefore the solution in the region
${\cal D}_2$ is parameterized by the asymptotic values of the data
for the region ${\cal D}_1$ corresponding to  the absence of cut-off
of the exponentially vanishing tail. The region ${\cal D}_2$
describes the phase when the slow-light soliton is stopped: the
fields vanish, while the information borne by the soliton is stored
in the medium in the form of spatially localized polarization. At
the time $T$ the laser is instantly turned on again. The stored
localized polarization then generates  a moving slow-light soliton.
This process is described by the solution in the region ${\cal
D}_3$. Except for the moment $T_1$, the functions $w,z$ are
continuous in $\tau$. This ensures  that the physical variables such
as the wave-function and field amplitudes evolve continuously.
\begin{figure}[thb]
\centerline{\includegraphics[width=70mm]{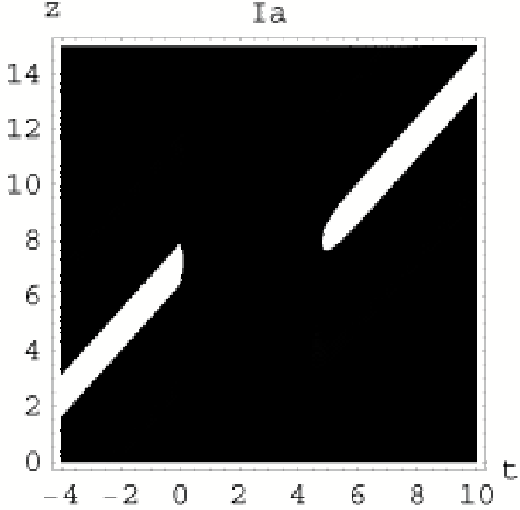}}
\caption{\label{fig2a} Contour plot of the intensity of
$\tilde\Omega_a$. We choose $\lambda=-4.1i$ and zero detuning,
$\Delta=0$. The break-up area in between the two solitonic trails
manifests the creation of a standing memory bit in the medium.}
\end{figure}

We demonstrate typical dynamics of the intensity of the fields in
Figs.\ref{fig1a} and \ref{fig2a}. In Fig.\ref{fig1a} the decaying
shock wave, whose front has an exponential profile, propagates with
the speed of light, reaches the slow-light soliton and stops it. An
intense and narrow peak is developing in the background field when
the shock wave hits the soliton (dotted curve). This peak signifies
a transfer of energy from the soliton to the background field. After
the auxiliary laser has been switched on again, another step-like
shock wave   reaches the localized polarization formed in the medium
by the incoming soliton, and retrieves stored information in the
form of a new slow-light soliton. A narrow and deep depression   in
the intensity plot means now the energy transfers in the opposite
direction, i.e. from the background field to the restored soliton.
The dynamics of the field $\tilde\Omega_a$ is plotted in
Fig.\ref{fig2a}. The contour plot shows that in the process of rapid
deceleration the solitonic trail profiles end sharply. Notice that
the characteristics of the restored pulse, i.e. the width and group
velocity, are very close to those of the input signal existing in
the medium before the stopping.

We now calculate the half-width of the polarization flip written
into the medium after the soliton is completely stopped. It reads
\begin{equation}\label{width_P}
{\cal W}_s= 4c\ln(2+\sqrt3)
    \frac{|\Delta-\lambda|^2}{\nu_0\left|\mathrm{Im}(\lambda)\right|}.
\end{equation}
It is important to notice that the width  Eq.(\ref{width_P}) of the
optical memory formation does not depend on the  rate $\alpha$. In
other words, the width of the memory bit is not sensitive to how
rapidly, i.e. nonadiabatically, the controlling field is switched
off. This leads to an important conclusion. Indeed, through the
variation of the experimentally adjustable parameter $\alpha$, it is
possible to control the location of the  memory bit, while its
characteristic size remains intact.  Dutton and Hau have already
reported~\cite{Dutton:2004} that when the switching is made quickly
compared to the natural lifetime of the upper level, the adiabatic
assumptions are no longer valid, but, remarkably, the quality of the
storage is not reduced in the nonadiabatic regime. Our analytical
result is in excellent agreement with this observation.

The group velocity of the slow-light soliton reads
\begin{equation}\label{speed1}
    \frac{v_g}c=\frac{|w(\tau,\lambda)|^2}
    {\frac{\nu_0(1+|w(\tau,\lambda)|^2)}{2|\Delta-\lambda|^2}+|w(\tau,\lambda)|^2}.
\end{equation}
Notice that in the case of the constant background field, i.e. in
the case $\alpha=0$, the conventional expressions for the slow-light
soliton Eqs.(\ref{sl_field_a}),(\ref{sl_field_b})  along with the
expression for the group velocity Eq.(\ref{ss_vg}) - the main
motivational quantity for this report - can be readily recovered
from Eqs.(\ref{fields_tilde}),(\ref{speed1}).

The distance ${\cal L}_s(\alpha)$ that the slow-light soliton will
propagate from the moment when the laser is switched off until the
full stop, is
\begin{eqnarray}\label{delta_L}
  &{\cal L}_s(\alpha)=\frac{2c|\Delta-\lambda|^2}{\nu_0\mathrm{Im}(\lambda)}
  \,\tilde\phi_s|_{\tau=0}^{\tau=\infty}\nonumber\\
  =& \frac{2c|\Delta-\lambda|^2}{\nu_0|\mathrm{Im}(\lambda)|}
\left[\ln\sqrt{1+|w_0|^2}-\mathrm{Re}(z(\infty,\lambda))\right].\quad
\end{eqnarray}
It is evident from our solution Eq.(\ref{delta_L}) that the
soliton possesses  some inertia or, in other words, a momentum of
motion. Indeed, even if the controlling field is  switched off
instantly (notice that
$\lim\limits_{\alpha\to\infty}\mathrm{Re}\,z(\infty,\lambda)=0$),
the soliton will still propagate over some finite distance  after
the shock wave of the vanishing field, propagating with the speed
of light, has reached the soliton.

\begin{figure}[thb]
\centerline{\includegraphics[width=80mm]{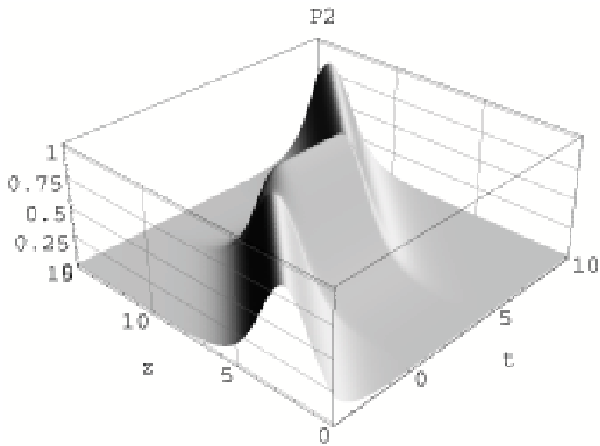}}
\caption{\label{fig3a} Population of level $|2\rangle$. Here,
$\lambda=-4.1i$ and $\Delta=0$. The  time interval $1 \le t \le 4$
corresponds to a standing localized  polarization flip. Compare with
experimental results in Fig.1 of ref.~\cite{Bajcsy:2003}.}
\end{figure}
In Fig.\ref{fig3a} we show the dynamics of  localized polarization
corresponding to soliton dynamics shown in Fig.\ref{fig2a}. The
plot is remarkably similar to the corresponding figure in
reference ~\cite{Bajcsy:2003} describing recent experiments. The
central part of the plot in Fig.\ref{fig3a} shows the standing
memory bit imprinted by the slow-light soliton. Notice that in the
presence of the soliton the population flip from level $|1\rangle$
to $|2\rangle$ in the center of the peak is almost complete. This
property of the solution manifests the major distinction between
the strongly nonlinear regime considered in our paper and the
linear EIT theory. As  was already pointed out earlier
\cite{Matsko:2001, Dey:2003, Dutton:2004}, the regime of two
fields being comparable in magnitude opens up new avenues for an
effective control over superposition of two lower states of the
atoms. Changing the parameters of the slow-light soliton we can
coherently drive the system to access any point on the Bloch
sphere, which describes the lower  levels. For zero detuning, the
solution discussed here shows that when the field vanishes the
maximum population of the second level reaches unity. Using
Eq.(\ref{atoms_tilde}) it is also not difficult to estimate the
maximum population of the level $|2\rangle$ for finite detuning
after the soliton was completely stopped: $
{|\lambda|^2}/{|\Delta-\lambda|^2}$.
Notice that only a small fraction of the total population is located
in the upper level $|3\rangle$ and provides for some atom-field
interaction. This population is proportional to $|\Omega_0|^2$.
Numerical studies of the Maxwell-Bloch equations with  relaxation
terms included \cite{Dey:2003, Dutton:2004} show that for
experimentally feasible group velocities of the slow-light pulses,
i.e. when the maximum intensity of the controlling field is not very
high, the pulses are stable against relaxation from level
$|3\rangle$. Here we consider the same range of parameters.
Therefore, the destructive influence of relaxation on our solution
Eq.(\ref{fields_tilde}) is negligible.

\section{\label{sec5:level1}The case of arbitrary controlling field}

In this section we  build a single-soliton solution on the
background of the state of the overall atom-field system described
by Eq.(\ref{init_fields0}) for a quit general (complex) controlling
field $\Omega(\tau)$. The single-soliton solution corresponding to
the background field $\Omega(\tau)$ is given by
Eqs.(\ref{fields_tilde}),(\ref{atoms_tilde}) with the functions
$w(\tau,\lambda)$ and $z(\tau,\lambda)$  defined below.

We envisage the dynamics scenario similar to that of the previous
section. We assume that the slow-light soliton   is propagating in
nonlinear superposition with the background field, which is
constant at $\tau\to-\infty$ and vanishes at $\tau\to +\infty$.
The speed of the slow-light soliton is controlled by the intensity
of the background field. Therefore, when the background field
decreases, the slow-light soliton  slows down and stops,
eventually disappearing and leaving behind a standing localized
polarization flip, i.e. optical memory bit. Should the background
field increase, the soliton will emerge again and accelerate
accordingly.

To be specific, we define the asymptotic behavior for the field
$\Omega(\tau)$ in the form
\begin{equation}\label{field_asym}
    \Omega(\tau\rightarrow-\infty)=\Omega_0,\;
    \Omega(\tau\rightarrow+\infty)=0.
\end{equation}
The  asymptotic boundary conditions Eq.(\ref{field_asym}) dictate
the following asymptotic behavior for the functions
$w(\tau,\lambda)$ and $z(\tau,\lambda)$ defined by Eqs.
(\ref{Riccati}),(\ref{Zequation}):
\begin{eqnarray}\label{w_asym}
w(-\infty,\lambda)&=& w_0 =\frac{\Omega_0}{2 k(\lambda)},\;
w(+\infty,\lambda)=0,\\
\label{z_asym}z(-\infty,\lambda)&=& z_0\tau=i\frac
{|\Omega_0|^2}{4k(\lambda)} \tau,
\end{eqnarray}
where $k(\lambda)=(\lambda+\sqrt{\lambda^2+|\Omega_0|^2})/2$. The
function $z(\tau,\lambda)$ satisfying  the asymptotical conditions
Eq.(\ref{z_asym}) reads
\begin{equation}\label{z_regul}
    z(\tau,\lambda)=z_0\tau+
\int\limits_{-\infty}^\infty\left({\frac i2\Omega^*(\tau')
w(\tau',\lambda)-z_0}\right)\Theta(\tau-\tau')d\tau'.
\end{equation}

The  function $w(\tau,\lambda)$  is defined by the relations

\begin{eqnarray}\label{w_solut}
w(\tau,\lambda)&=&i\int\limits_{-\infty}^\infty e^{-i\,k(\tau-s)}
\Theta(\tau-s) \tilde w(s,\lambda)\,ds,\\
 \label{Riccati_asym}
\tilde w(\tau,\lambda)&=&\frac{\Omega(\tau)}2 +\frac1{k^2}
\left({\frac{|\Omega_0|^2}4 k\,w-\frac{\Omega^*(\tau)}2
(k\,w)^2}\right)\!.\quad\end{eqnarray} Here $\Theta(\tau)$ is the
Heaviside step function. We rewrite the relations
Eqs.(\ref{w_solut}),(\ref{Riccati_asym}) in the form of nonlinear
integral equation, viz.
\begin{eqnarray}\label{Riccati_asym1}
&\tilde w(\tau,\lambda)=\frac{\Omega(\tau)}2
+\int\limits_{-\infty}^\infty
e^{-i\,k(\tau-s)} \Theta(\tau-s) \tilde w(s,\lambda)\,ds\nonumber\\
&\cdot\int\limits_{-\infty}^\infty e^{-i\,k(\tau-s)}
\Theta(\tau-s) \left({\frac{\Omega^*(\tau)}2 \tilde
w(s,\lambda)-\frac{|\Omega_0|^2}4 }\right)\,ds.\quad\end{eqnarray}
\noindent Hence, we can construct a solution  $\tilde
w(\tau,\lambda)$ iterating Eq.(\ref{Riccati_asym1}) and starting
iterations from $\tilde w_0(\tau,\lambda)=\frac12\Omega(\tau)$.
 \noindent Notice that the last  term in Eq.(\ref{Riccati_asym})
provides a correction of order $k^{-2}$, because the function
$w(\tau,\lambda)$ asymptotically behaves as $1/k$. In the adiabatic
regime when the background field varies slowly, i.e. all derivatives
of  $\Omega(\tau)$ are much smaller than $k$, we can integrate
Eq.(\ref{w_solut}) by parts. Preserving only the lowest order term
with respect to $k$ we obtain
\begin{equation}\label{w_first}
    w(\tau,\lambda)\approx \frac{\Omega(\tau)}{2k}.
\end{equation}
Hence, at the lowest order in $k$ we find
\begin{equation}\label{z_first}
     z(\tau,\lambda)\approx z_0\tau+
\int\limits_{-\infty}^\infty\left({\frac i{4k}|\Omega(\tau')|^2
-z_0}\right)\Theta(\tau-\tau')d\tau'.
\end{equation}
As one can easily observe this expression is in agreement with the
asymptotic condition Eq.(\ref{z_asym}).

For an arbitrary dependence of the background field on the retarded
time $\tau$, the speed of the slow-light soliton can be represented
in the following
 form:
\begin{equation}\label{speed1_1}
    \frac{v_g}c=\frac{\partial_\tau\, \phi_s}
    {\partial_\tau\, \phi_s-\partial_\zeta\, \phi_s}.
\end{equation}
It can be readily seen that
\begin{equation}\label{dtau_phi}
\frac{\partial\phi_s}{\partial\tau}=\frac{\mathrm{Im}(\lambda)
|w(\tau,\lambda)|^2}{1+|w^(\tau,\lambda)|^2},\;
\frac{\partial\phi_s}{\partial\zeta}=\frac{\nu_0}{2}
\mathrm{Im}\frac1{\lambda-\Delta}.
\end{equation}
We have thus found a general solution for the velocity $v_g$ of the
slow-light soliton propagating on   an arbitrary time-dependent
background field in terms of the function $\tilde w(\tau,\lambda)$
given by Eq.(\ref{Riccati_asym1}). This result provides a new way to
study dynamics of localized optical signals in the nonlinear EIT
systems. It allows us to easily suggest different schemes to
slow-light down, stop, and reaccelerate slow-light solitonic
contribution in the probing pulse. With such techniques one can
introduce a concept of probing different regions of the media by
changing the time that the soliton dwells at a particular location.
This time is important in situations where the interaction between
light and some impurities inside the EIT medium is weak and requires
slowing the signal down in the vicinity of these impurities in order
to gain more information about the structure of the medium.

We also introduce a notion of the distance ${\cal L}[\Omega]$ that
the slow-light soliton will propagate until it fully stops. This
quantity is important because it describes the location of an
imprinted memory bit. The brackets $[\cdot]$ indicate a functional
dependence of the distance on the controlling field
$\Omega(\tau)$. To begin with we consider the case when the field
is  instantly switched off at the moment $\tau=0$, i.e.
$\Omega(\tau)=\Omega_0\Theta(-\tau)$. Then we easily find the
solution for $w$ and $z$:
$$w(\tau,\lambda)=w_0\left({\Theta(-\tau)+\Theta(\tau)\;
e^{-i\lambda\tau}}\right),\;z(\tau,\lambda)=z_0\Theta(-\tau)\tau.$$
Hence, we can obtain the distance ${\cal L}_0$ that the soliton will
propagate from the moment $\tau=0$ until its complete stop at
$\tau\rightarrow\infty$:
$$
{\cal L}_0=\frac{c|\Delta-\lambda|^2}{\nu_0|\mathrm{Im}(\lambda)|}
\ln\left(1+|w_0|^2\right).
$$
Here we make use of the assumption that $\mathrm{Im}(\lambda)<0$.

Now, we can give the definition of the distance ${\cal L}[\Omega]$
for a generic field $\Omega(\tau)$ satisfying the conditions
Eq.(\ref{field_asym}). It is convenient to define it as a relative
distance, namely the difference between the absolute coordinate of
the stopped signal at the maximum of the signal and the distance
${\cal L}_0$. The relative distance reads:
\begin{eqnarray}\label{delta_L_1}
  {\cal L}[\Omega]=\frac{2c|\Delta-\lambda|^2}{\nu_0\mathrm{Im}(\lambda)}
 \!\!\int\limits_{-\infty}^\infty  \mathrm{Re}\left({\frac
i2\Omega^*(\tau)
w(\tau,\lambda)-z_0\Theta(-\tau)}\right)d\tau.\nonumber
\end{eqnarray}
Using the representation Eq.(\ref{Riccati_asym}) we find
\begin{widetext}
\begin{eqnarray}\label{delta_L_asym}
{\cal L}[\Omega]=
\frac{2c|\Delta-\lambda|^2}{\nu_0\mathrm{Im}(\lambda)}
\mathrm{Re}\left({\int\limits_{-\infty}^{+\infty}
\int\limits_{-\infty}^{+\infty} e^{-i\,k(\tau-s)}
\Theta(\tau-s)\left({\frac{|\Omega_0|^2}4\Theta(-\tau)-\frac{\Omega^*(\tau)}2
\tilde w(s,\lambda)}\right)ds\,d\tau}\right)_.
\end{eqnarray}
\end{widetext}
If we assume that $\Omega(\tau)$ is a smooth function and
substitute the solution for $\tilde w(\tau,\lambda)$, we find the
result in the form of a series
$${\cal L}[\Omega]=\frac{2c|\Delta-\lambda|^2}{2\nu_0\mathrm{Im}(\lambda)}
\mathrm{Im}\left({\sum_{n=1}^\infty \frac{I_n}{k^n}}\right),$$ where
$I_n[\Omega]$ are regularized Zakharov-Shabat
functionals~\cite{fad}. The first two functionals read
$I_1[\Omega]=-\int_{-\infty}^{\infty}\left({|\Omega(\tau)|^2-|\Omega_0|^2\Theta(-\tau)}
\right) d\tau$,
$I_2[\Omega]=\frac1{2i}\int_{-\infty}^{\infty}(\Omega^*(s)\partial_s
\Omega(s)-\Omega(s)\partial_s \Omega^*(s)) ds$. The other
functionals can be obtained through the iteration procedure
described above. As it is usual for the boundary conditions of
finite density type,  $I_1$ is not a proper functional on the
complex manifold  of  physical observables, in the sense described
in~\cite{fad}. In that sense all other functionals in the expansion
with respect to $k$ are proper. It is a  plausible conjecture that
the minimum of the functional of length, Eq.(\ref{delta_L_asym}),
i.e. $\delta{\cal L}[\Omega]/\delta\Omega=0$ with $\delta^2{\cal
L}[\Omega]/\delta\Omega^2>0$, is achieved when the controlling field
is switched off instantly. Therefore it seems intuitively correct
that the minimum is delivered by the the function
$\Omega_0\Theta(-\tau)$ discussed above. This conjecture is also
supported by the case discussed in section~\ref{sec4:level1}, when
the controlling field vanishes exponentially, i.e.
$\Omega(\tau)=\Omega_0(\Theta(-\tau)+\Theta(\tau)e^{-\alpha\tau})$.
In this case the minimum of length is delivered by a singular limit
$\alpha\to\infty$, i.e. in the regime of instant switching off of
the controlling field.

Another important characteristics of the system is the shape of the
imprinted signal. It is easy to show that the width ${\cal W}_0$ of
the imprinted memory bit is not sensitive to the functional form of
$\Omega(\tau)$ and is the same as given by Eq.(\ref{width_P}). In
other words, this exact result is valid regardless of how rapidly we
switch the background field off. This means that specification of
$\Omega(\tau)$ only influences the location of the stored signal and
not its shape. This result is strongly supported by recent
experiments \cite{Dutton:2004}. This reference emphasizes the
phenomenological fact that the quality of the storage is not
sensitive to the regime of  switching off of the control laser. Our
exact result Eq.(\ref{width_P}) provides a rigorous basis for this
experimental observation.

\begin{figure}[thb]
\centerline{\includegraphics[width=50mm]{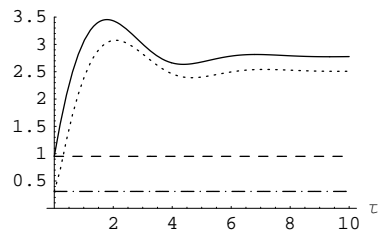}}
\caption{\label{fig1b} Here $\varepsilon_0=4.1$, $\Delta=0$, and
$\Omega(\tau)=0.5 e^{-4\tau}$. It is clear that when the background
field changes relatively fast the exact solution given by solid and
dotted lines are different from approximative one (dashed and
dot-and-dash lines, correspondingly) by up to $40\%$.}
\end{figure}

To conclude this section we  discuss the applicability of the
concept of effective time to the regime of nonadiabatic variations
of the controlling field. This concept was used before in
\cite{Grobe:1994, Bajcsy:2003, Matsko:2001} for approximative
descriptions of pulse propagation on  the background of a
time-dependent controlling field. To account for this dependence,
these references introduce an effective time variable (a "scaled
time" of ref.\cite{Bajcsy:2003} and see the function $Z(\tau)$ of
ref. \cite{Grobe:1994} ). The reference \cite{Grobe:1994}) shows the
concept of effective (scaled) time to be very useful in the regime
of linear EIT, while ref. \cite{Bajcsy:2003} demonstrates
applicability of this concept in the strongly nonlinear, though
adiabatic, regime. The effective time approximation, in our
formulation, is given by expressions
Eqs.(\ref{w_first}),(\ref{z_first}) employed in the slow-light
soliton solution Eqs.(\ref{fields_tilde}),(\ref{atoms_tilde}). In
Fig.~\ref{fig1b} we compare the resulting approximate solution
against an exactly solvable case discussed in
section~\ref{sec4:level1}. We point out that the method of effective
time has a rather limited applicability in the  strongly nonlinear
noadiabatic regime, which is most interesting for modern
experiments~\cite{Dutton:2004}. Indeed, Fig.~\ref{fig1b}
demonstrates that this method as applied to the slow-light soliton
gives rise to large errors in the regime of nonadiabatic dynamics.
Therefore heuristic attempts~\cite{ulf:2004} to substitute an
effective time into the slow-light soliton solution
Eqs.(\ref{sl_field_a}),(\ref{sl_field_b}) is largely not  accurate.

\section{\label{sec6:level1} Conclusions and discussion }
In this paper we have discussed a physically realistic, exactly
solvable model of manipulation, i.e. preparation, control and
readout, of optical memory bits in the strongly nonlinear, and more
importantly, nonadiabatic regime. We discussed a concept of the
transparency gate for the  slow-light solitons in a $\Lambda$-type
medium. We explained how the fast soliton can destroy the slow-light
soliton and close the gate for the latter. We
 also described the process of reading optical information,
written into the active medium by the slow-light solitons. We have
investigated a mechanism of dynamical control  of the slow-light
soliton whose group velocity explicitly depends on the background
field. For a quite general background field, we found the location
and shape of the memory bit written into the medium upon stopping
the signal. Remarkably, the width of this spatially localized
standing polarization flip is not sensitive at all to the functional
form of the controlling field and is defined by the parameters of
the slow-light soliton only.

It is worth discussing here a possibility to actually create in the
$\Lambda$-type atomic medium the  slow-light solitons. The general
physical feature underlying the mathematical property of complete
integrability is a  delicate balance between the effects of
dispersion and nonlinearity  inherent in the medium. Provided that
this balance is observed and the system is completely integrable, it
is a general fact that virtually any sufficiently intensive
localized initial condition creates solitons. The overall picture of
nonlinear dynamics can be roughly described as follows. The evolving
signal created by the incident pulse in the course of the dynamics
separates into a number of solitons and a decaying tail. The latter
vanishes in due course. The soliton-like signals survive (ideally,
i.e. in the absence of dissipation) for infinitely long time.   When
the physical conditions underlying the complete integrability of the
optical  system are met, the general picture of the nonlinear
dynamics is  similar to that described above. Namely, a fairly
arbitrary localized  and intensive initial signal creates in the
course of nonlinear dynamics a number of slow-light solitons. The
number of  slow-light solitons can be derived from the analysis of
the corresponding zero-curvature representation. The signal
Eqs.(\ref{sl_field_a}),(\ref{sl_field_b}) is a  generic soliton-like
solution of the nonlinear system Eqs.(\ref{Maxwell_m}),
(\ref{Liouv}) and therefore it is very plausible that such signal
can be created. To conclude, we want to emphasize that in our
considerations the distinguished role  is assigned to the background
field $\Omega(\tau)$ that turns out to be a nonlinear analog of the
conventional controlling field appearing in the linear EIT
formulation. The difference between the linear and nonlinear cases
lies in the fact that in the nonlinear case the control field and
the slow-light soliton solution are present in the {\it same}
channel in an inseparable nonlinear superposition.

\section{\label{sec7:level1}Acknowledgements}

IV acknowledges the support of the Engineering and Physical Sciences
Research Council, United Kingdom. Work at Los Alamos National
Laboratory is supported by the USDoE.


\begin{thebibliography}{31}
\expandafter\ifx\csname
natexlab\endcsname\relax\def\natexlab#1{#1}\fi
\expandafter\ifx\csname bibnamefont\endcsname\relax
  \def\bibnamefont#1{#1}\fi
\expandafter\ifx\csname bibfnamefont\endcsname\relax
  \def\bibfnamefont#1{#1}\fi
\expandafter\ifx\csname citenamefont\endcsname\relax
  \def\citenamefont#1{#1}\fi
\expandafter\ifx\csname url\endcsname\relax
  \def\url#1{\texttt{#1}}\fi
\expandafter\ifx\csname urlprefix\endcsname\relax\def\urlprefix{URL
}\fi \providecommand{\bibinfo}[2]{#2}
\providecommand{\eprint}[2][]{\url{#2}}

\bibitem[{\citenamefont{Hau et~al.}(1999)\citenamefont{Hau, Harris, Dutton, and
  Behroozi}}]{Hau:1999}
\bibinfo{author}{\bibfnamefont{L.~N.} \bibnamefont{Hau}},
  \bibinfo{author}{\bibfnamefont{S.~E.} \bibnamefont{Harris}},
  \bibinfo{author}{\bibfnamefont{Z.}~\bibnamefont{Dutton}}, \bibnamefont{and}
  \bibinfo{author}{\bibfnamefont{C.~H.} \bibnamefont{Behroozi}},
  \bibinfo{journal}{Lett.\ to Nature} \textbf{\bibinfo{volume}{397}},
  \bibinfo{pages}{594} (\bibinfo{year}{1999}).

\bibitem[{\citenamefont{Liu et~al.}(2001)\citenamefont{Liu, Dutton, Behroozi,
  and Hau}}]{Liu:2001}
\bibinfo{author}{\bibfnamefont{C.}~\bibnamefont{Liu}},
  \bibinfo{author}{\bibfnamefont{Z.}~\bibnamefont{Dutton}},
  \bibinfo{author}{\bibfnamefont{C.~H.} \bibnamefont{Behroozi}},
  \bibnamefont{and} \bibinfo{author}{\bibfnamefont{L.~V.} \bibnamefont{Hau}},
  \bibinfo{journal}{Lett.\ to Nature} \textbf{\bibinfo{volume}{409}},
  \bibinfo{pages}{490} (\bibinfo{year}{2001}).

\bibitem[{\citenamefont{Phillips et~al.}(2001)\citenamefont{Phillips,
  Fleischhauer, Mair, Walsworth, and Lukin}}]{Phillips:2001}
\bibinfo{author}{\bibfnamefont{D.~F.} \bibnamefont{Phillips}},
  \bibinfo{author}{\bibfnamefont{A.}~\bibnamefont{Fleischhauer}},
  \bibinfo{author}{\bibfnamefont{A.}~\bibnamefont{Mair}},
  \bibinfo{author}{\bibfnamefont{R.~L.} \bibnamefont{Walsworth}},
  \bibnamefont{and} \bibinfo{author}{\bibfnamefont{M.~D.} \bibnamefont{Lukin}},
  \bibinfo{journal}{Phys. Rev. Lett.} \textbf{\bibinfo{volume}{86}},
  \bibinfo{pages}{783} (\bibinfo{year}{2001}).

\bibitem[{\citenamefont{Bajcsy et~al.}(2003)\citenamefont{Bajcsy, Zibrov, and
  Lukin}}]{Bajcsy:2003}
\bibinfo{author}{\bibfnamefont{M.}~\bibnamefont{Bajcsy}},
  \bibinfo{author}{\bibfnamefont{A.~S.} \bibnamefont{Zibrov}},
  \bibnamefont{and} \bibinfo{author}{\bibfnamefont{M.~D.} \bibnamefont{Lukin}},
  \bibinfo{journal}{Lett.\ to Nature} \textbf{\bibinfo{volume}{426}},
  \bibinfo{pages}{638} (\bibinfo{year}{2003}).

\bibitem[{\citenamefont{Braje et~al.}(2003)\citenamefont{Braje, Balic, Yin, and
  Harris}}]{Braje:2003}
\bibinfo{author}{\bibfnamefont{D.~A.} \bibnamefont{Braje}},
  \bibinfo{author}{\bibfnamefont{V.}~\bibnamefont{Balic}},
  \bibinfo{author}{\bibfnamefont{G.~Y.} \bibnamefont{Yin}}, \bibnamefont{and}
  \bibinfo{author}{\bibfnamefont{S.~E.} \bibnamefont{Harris}},
  \bibinfo{journal}{Phys. Rev. A} \textbf{\bibinfo{volume}{68}},
  \bibinfo{pages}{041801(R)} (\bibinfo{year}{2003}).

\bibitem[{\citenamefont{Mikhailov et~al.}(2004)\citenamefont{Mikhailov,
  Sautenkov, Rostovtsev, and Welch}}]{Mikhailov:2004}
\bibinfo{author}{\bibfnamefont{E.~E.} \bibnamefont{Mikhailov}},
  \bibinfo{author}{\bibfnamefont{V.~A.} \bibnamefont{Sautenkov}},
  \bibinfo{author}{\bibfnamefont{Y.~V.} \bibnamefont{Rostovtsev}},
  \bibnamefont{and} \bibinfo{author}{\bibfnamefont{G.~R.} \bibnamefont{Welch}},
  \bibinfo{journal}{J. Opt. Soc. Am. B} \textbf{\bibinfo{volume}{21}},
  \bibinfo{pages}{425} (\bibinfo{year}{2004}).

\bibitem[{\citenamefont{Turukhin et~al.}(2002)\citenamefont{Turukhin,
  Sudarshanam, Shahriar, and et~al.}}]{Turukhin:2002}
\bibinfo{author}{\bibfnamefont{A.~V.} \bibnamefont{Turukhin}},
  \bibinfo{author}{\bibfnamefont{V.~S.} \bibnamefont{Sudarshanam}},
  \bibinfo{author}{\bibfnamefont{M.~S.} \bibnamefont{Shahriar}},
  \bibnamefont{and} \bibinfo{author}{\bibnamefont{et~al.}},
  \bibinfo{journal}{Phys.\ Rev.\ Lett.} \textbf{\bibinfo{volume}{88}},
  \bibinfo{pages}{023602} (\bibinfo{year}{2002}).

\bibitem[{\citenamefont{Bigelow et~al.}(2003)\citenamefont{Bigelow, Lepeshkin,
  and Boyd}}]{Bigelow:2003}
\bibinfo{author}{\bibfnamefont{M.~S.} \bibnamefont{Bigelow}},
  \bibinfo{author}{\bibfnamefont{N.~N.} \bibnamefont{Lepeshkin}},
  \bibnamefont{and} \bibinfo{author}{\bibfnamefont{R.~W.} \bibnamefont{Boyd}},
  \bibinfo{journal}{Science} \textbf{\bibinfo{volume}{301}},
  \bibinfo{pages}{200} (\bibinfo{year}{2003}).

\bibitem[{\citenamefont{Soljacic and Joannopoulos}(2004)}]{Soljacic:2004}
\bibinfo{author}{\bibfnamefont{M.}~\bibnamefont{Soljacic}} \bibnamefont{and}
  \bibinfo{author}{\bibfnamefont{J.~D.} \bibnamefont{Joannopoulos}},
  \bibinfo{journal}{Nature Materials} \textbf{\bibinfo{volume}{3}},
  \bibinfo{pages}{213} (\bibinfo{year}{2004}).

\bibitem[{\citenamefont{Kocharovskaya et~al.}(2001)\citenamefont{Kocharovskaya,
  Rostovtsev, and Scully}}]{Kocharovskaya:2001}
\bibinfo{author}{\bibfnamefont{O.}~\bibnamefont{Kocharovskaya}},
  \bibinfo{author}{\bibfnamefont{Y.}~\bibnamefont{Rostovtsev}},
  \bibnamefont{and} \bibinfo{author}{\bibfnamefont{M.~O.}
  \bibnamefont{Scully}}, \bibinfo{journal}{Phys. Rev. Lett.}
  \textbf{\bibinfo{volume}{86}}, \bibinfo{pages}{628} (\bibinfo{year}{2001}).

\bibitem[{\citenamefont{Dutton and Hau}(2004)}]{Dutton:2004}
\bibinfo{author}{\bibfnamefont{Z.}~\bibnamefont{Dutton}} \bibnamefont{and}
  \bibinfo{author}{\bibfnamefont{L.~V.} \bibnamefont{Hau}},
  \bibinfo{journal}{Phys. Rev. A} \textbf{\bibinfo{volume}{70}},
  \bibinfo{pages}{053831} (\bibinfo{year}{2004}).

\bibitem[{\citenamefont{Harris}(1997)}]{Harris:1997}
\bibinfo{author}{\bibfnamefont{S.~E.} \bibnamefont{Harris}},
  \bibinfo{journal}{Phys. Today} \textbf{\bibinfo{volume}{50(7)}},
  \bibinfo{pages}{36} (\bibinfo{year}{1997}).

\bibitem[{\citenamefont{Lukin}(2003)}]{Lukin:2003}
\bibinfo{author}{\bibfnamefont{M.~D.} \bibnamefont{Lukin}},
  \bibinfo{journal}{Rev. Mod. Phys.} \textbf{\bibinfo{volume}{75}},
  \bibinfo{pages}{457} (\bibinfo{year}{2003}).

\bibitem[{\citenamefont{Grobe et~al.}(1994)\citenamefont{Grobe, Hioe, and
  Eberly}}]{Grobe:1994}
\bibinfo{author}{\bibfnamefont{R.}~\bibnamefont{Grobe}},
  \bibinfo{author}{\bibfnamefont{F.~T.} \bibnamefont{Hioe}}, \bibnamefont{and}
  \bibinfo{author}{\bibfnamefont{J.~H.} \bibnamefont{Eberly}},
  \bibinfo{journal}{Phys.\ Rev.\ Lett.} \textbf{\bibinfo{volume}{73}},
  \bibinfo{pages}{3183} (\bibinfo{year}{1994}).

\bibitem[{\citenamefont{Eberly}(1995)}]{Eberly:1995}
\bibinfo{author}{\bibfnamefont{J.~H.} \bibnamefont{Eberly}},
  \bibinfo{journal}{Quant. Semiclass. Opt.} \textbf{\bibinfo{volume}{7}},
  \bibinfo{pages}{373} (\bibinfo{year}{1995}).

\bibitem[{\citenamefont{Andreev}(1998)}]{andreev:1998}
\bibinfo{author}{\bibfnamefont{A.~V.} \bibnamefont{Andreev}},
  \bibinfo{journal}{JETP} \textbf{\bibinfo{volume}{86}}, \bibinfo{pages}{412}
  (\bibinfo{year}{1998}).

\bibitem[{\citenamefont{Dey and Agarwal}(2003)}]{Dey:2003}
\bibinfo{author}{\bibfnamefont{T.~N.} \bibnamefont{Dey}} \bibnamefont{and}
  \bibinfo{author}{\bibfnamefont{G.~S.} \bibnamefont{Agarwal}},
  \bibinfo{journal}{Phys. Rev. A} \textbf{\bibinfo{volume}{67}},
  \bibinfo{pages}{033813} (\bibinfo{year}{2003}).

\bibitem[{\citenamefont{Kozlov and Eberly}(2000)}]{Kozlov:2000}
\bibinfo{author}{\bibfnamefont{V.~V.} \bibnamefont{Kozlov}} \bibnamefont{and}
  \bibinfo{author}{\bibfnamefont{J.~H.} \bibnamefont{Eberly}},
  \bibinfo{journal}{Opt. Commun.} \textbf{\bibinfo{volume}{179}},
  \bibinfo{pages}{85} (\bibinfo{year}{2000}).

\bibitem[{\citenamefont{Dutton et~al.}(2001)\citenamefont{Dutton, Budde, Slowe,
  and Hau}}]{Dutton:2001}
\bibinfo{author}{\bibfnamefont{Z.}~\bibnamefont{Dutton}},
  \bibinfo{author}{\bibfnamefont{M.}~\bibnamefont{Budde}},
  \bibinfo{author}{\bibfnamefont{C.}~\bibnamefont{Slowe}}, \bibnamefont{and}
  \bibinfo{author}{\bibfnamefont{L.~V.} \bibnamefont{Hau}},
  \bibinfo{journal}{Science} \textbf{\bibinfo{volume}{293}},
  \bibinfo{pages}{663} (\bibinfo{year}{2001}).

\bibitem[{\citenamefont{Matsko et~al.}(2001)\citenamefont{Matsko, Rostovstsev,
  Kocharovskaya, Zibrov, and Scully}}]{Matsko:2001}
\bibinfo{author}{\bibfnamefont{A.~B.} \bibnamefont{Matsko}},
  \bibinfo{author}{\bibfnamefont{Y.~V.} \bibnamefont{Rostovstsev}},
  \bibinfo{author}{\bibfnamefont{O.}~\bibnamefont{Kocharovskaya}},
  \bibinfo{author}{\bibfnamefont{A.~S.} \bibnamefont{Zibrov}},
  \bibnamefont{and} \bibinfo{author}{\bibfnamefont{M.~O.}
  \bibnamefont{Scully}}, \bibinfo{journal}{Phys. Rev. A}
  \textbf{\bibinfo{volume}{64}}, \bibinfo{pages}{043809}
  (\bibinfo{year}{2001}).

\bibitem[{\citenamefont{Faddeev and Takhtadjan}(1987)}]{fad}
\bibinfo{author}{\bibfnamefont{L.~D.} \bibnamefont{Faddeev}} \bibnamefont{and}
  \bibinfo{author}{\bibfnamefont{L.~A.} \bibnamefont{Takhtadjan}},
  \emph{\bibinfo{title}{Hamiltonian Methods in the Theory of Solitons}}
  (\bibinfo{publisher}{Springer, Berlin}, \bibinfo{year}{1987}).

\bibitem[{\citenamefont{Park and Shin}(1998)}]{Park:1998}
\bibinfo{author}{\bibfnamefont{Q.-H.} \bibnamefont{Park}} \bibnamefont{and}
  \bibinfo{author}{\bibfnamefont{H.~J.} \bibnamefont{Shin}},
  \bibinfo{journal}{Phys.\ Rev.\ A} \textbf{\bibinfo{volume}{57}},
  \bibinfo{pages}{4643} (\bibinfo{year}{1998}).

\bibitem[{\citenamefont{Byrne et~al.}(2003)\citenamefont{Byrne, Gabitov, and
  Kova\v{c}i\v{c}}}]{gab}
\bibinfo{author}{\bibfnamefont{J.~A.} \bibnamefont{Byrne}},
  \bibinfo{author}{\bibfnamefont{I.~R.} \bibnamefont{Gabitov}},
  \bibnamefont{and}
  \bibinfo{author}{\bibfnamefont{G.}~\bibnamefont{Kova\v{c}i\v{c}}},
  \bibinfo{journal}{Physica D} \textbf{\bibinfo{volume}{186}},
  \bibinfo{pages}{69} (\bibinfo{year}{2003}).

\bibitem[{\citenamefont{Rybin and Vadeiko}(2004)}]{Rybin:2004}
\bibinfo{author}{\bibfnamefont{A.~V.} \bibnamefont{Rybin}} \bibnamefont{and}
  \bibinfo{author}{\bibfnamefont{I.~P.} \bibnamefont{Vadeiko}},
  \bibinfo{journal}{Journal of Optics B: Quantum and Semiclassical Optics}
  \textbf{\bibinfo{volume}{6}}, \bibinfo{pages}{416} (\bibinfo{year}{2004}).

\bibitem[{\citenamefont{Hioe and Grobe}(1994)}]{Hioe:1994}
\bibinfo{author}{\bibfnamefont{F.~T.} \bibnamefont{Hioe}} \bibnamefont{and}
  \bibinfo{author}{\bibfnamefont{R.}~\bibnamefont{Grobe}},
  \bibinfo{journal}{Phys.\ Rev.\ Lett.} \textbf{\bibinfo{volume}{73}},
  \bibinfo{pages}{2559} (\bibinfo{year}{1994}).

\bibitem[{\citenamefont{Rybin et~al.}(1988)\citenamefont{Rybin, Matveev, and
  Salle}}]{ryb1}
\bibinfo{author}{\bibfnamefont{A.~V.} \bibnamefont{Rybin}},
  \bibinfo{author}{\bibfnamefont{V.~B.} \bibnamefont{Matveev}},
  \bibnamefont{and} \bibinfo{author}{\bibfnamefont{M.~A.} \bibnamefont{Salle}},
  \bibinfo{journal}{Inverse Problems.} \textbf{\bibinfo{volume}{4}},
  \bibinfo{pages}{173} (\bibinfo{year}{1988}).

\bibitem[{\citenamefont{Rybin}(1991)}]{ryb2}
\bibinfo{author}{\bibfnamefont{A.~V.} \bibnamefont{Rybin}},
  \bibinfo{journal}{J. Phys. A: Math. and Gen.} \textbf{\bibinfo{volume}{24}},
  \bibinfo{pages}{5235} (\bibinfo{year}{1991}).

\bibitem[{\citenamefont{Rybin and Timonen}(1993)}]{ryb3}
\bibinfo{author}{\bibfnamefont{A.~V.} \bibnamefont{Rybin}} \bibnamefont{and}
  \bibinfo{author}{\bibfnamefont{J.}~\bibnamefont{Timonen}},
  \bibinfo{journal}{J. Phys. A: Math. And Gen.} \textbf{\bibinfo{volume}{26}},
  \bibinfo{pages}{3869} (\bibinfo{year}{1993}).

\bibitem[{\citenamefont{Matveev and Salle}(1991)}]{salle}
\bibinfo{author}{\bibfnamefont{V.~B.} \bibnamefont{Matveev}} \bibnamefont{and}
  \bibinfo{author}{\bibfnamefont{M.~A.} \bibnamefont{Salle}},
  \emph{\bibinfo{title}{Darboux Transformations and Solitons in Springer Series
  in Nonlinear Dynamics}} (\bibinfo{publisher}{Springer Verlag},
  \bibinfo{year}{1991}).

\bibitem[{\citenamefont{Fleischhauer and Lukin}(2000)}]{Fleischhauer:2000}
\bibinfo{author}{\bibfnamefont{M.}~\bibnamefont{Fleischhauer}}
  \bibnamefont{and} \bibinfo{author}{\bibfnamefont{M.~D.} \bibnamefont{Lukin}},
  \bibinfo{journal}{Phys.\ Rev.\ Lett.} \textbf{\bibinfo{volume}{84}},
  \bibinfo{pages}{5094} (\bibinfo{year}{2000}).

\bibitem[{\citenamefont{Leonhardt}(2004)}]{ulf:2004}
\bibinfo{author}{\bibfnamefont{U.}~\bibnamefont{Leonhardt}},
  \bibinfo{journal}{Preprint ArXiv: quant-ph/0408046}  (\bibinfo{year}{2004}).

\end{thebibliography}
\end{document}